\documentclass[showpacs,amssymb,amsmath,aps,pre,reprint,groupedaddress,nofootinbib,floatfix]{revtex4-2}
\usepackage[english]{babel}
\usepackage{graphicx}
\usepackage{bm}
\usepackage[usenames, dvipsnames]{xcolor}

\definecolor{blue(ryb)}{rgb}{0.01, 0.28, 1.0}
\usepackage{hyperref}
\hypersetup{citecolor=blue(ryb), linkcolor=blue(ryb), urlcolor=blue(ryb), anchorcolor=red, colorlinks=true}
\usepackage{orcidlink}

\newcommand{\br}{\mathbf{r}}
\newcommand{\bG}{\mathbf{G}}
\newcommand{\bg}{\mathbf{g}}
\newcommand{\nh}{h}
\newcommand{\e}{\mathrm{e}}
\newcommand{\nb}{\bar{n}}
\DeclareMathOperator{\Sp}{Sp}

\begin{document}
\title{Nonequilibrium protection effect and spatial localization of noise-induced fluctuations: quasi-one-dimensional driven lattice gas with partially penetrable obstacle
}

\author{S. P. Lukyanets\,\orcidlink{0000-0003-1169-2826}}
\email{lukyan[at]iop.kiev.ua}
\author{O. V. Kliushnichenko\,\orcidlink{0000-0003-4806-8971}}
\email{kliushnychenko[at]iop.kiev.ua}
\affiliation{Department of Theoretical Physics, Institute of Physics, NAS of Ukraine, Prospect Nauky 46, 03028 Kyiv, Ukraine
}

\begin{abstract}
We consider a nonequilibrium transition that leads to the formation of nonlinear steady-state structures due to the gas flow scattering on a partially penetrable obstacle. The resulting nonequilibrium steady state (NESS) corresponds to a two-domain gas structure attained at certain critical parameters.
We use a simple mean-field model of the driven lattice gas with ring topology to demonstrate that this transition is accompanied by the emergence of local invariants related to a complex composed of the obstacle and its nearest gas surrounding, which we refer to as obstacle edges. These invariants are independent of the main system parameters and behave as local first integrals, at least qualitatively.
As a result, the complex becomes insensitive to the noise of external driving field within the overcritical domain. The emerged invariants describe the conservation of the number of particles inside the obstacle and  strong temporal synchronization or correlation of gas states at obstacle edges. Such synchronization guarantees the equality to zero of the total edge current at any time. The robustness against external drive fluctuations is shown to be accompanied by strong spatial localization of induced gas fluctuations near the domain wall separating the depleted and dense gas phases.
Such a behavior can be associated with nonequilibrium protection effect and  synchronization of edges. The transition rates between different NESSs are shown to be different. The relaxation rates from one NESS to another take complex and real values in the sub- and overcritical regimes, respectively. The mechanism of these transitions is governed by  the generation of shock waves at the back side of the obstacle. In the subcritical regime, these solitary waves are generated sequentially many times, while only a single excitation is sufficient to rearrange the system state in the overcritical regime.
\end{abstract}
\maketitle

\textit{Note added.}---An erratum correcting typographical errors in this paper
has been published in Phys.~Rev.~E \textbf{113}, 049905(E) (2026),
\href{https://doi.org/10.1103/prqh-m74x}{DOI:~10.1103/prqh-m74x}.

\section{Introduction}

The evolution of many dissipative or out-of-equilibrium systems results in the formation of  steady states, in particular NESSs, implying the system decomposition into parts. Some of these parts or some of the resulting states may become insensitive to the external noise, fluctuations, or perturbations in the system. Often, such a decomposition is related to topological phenomena, e.g., non-Hermitian skin effect, edge correlation effect, topological or symmetry protection effects \cite{shankar2022,bal2019,hatano1996,Torres2018,yao2018,torres2019perspective,song2019,okuma2020,ashida2020,mori2020,wanjura2020, kawabata2023,garbe2023,minoguchi2023,cao2021,hu2022,zhang2022, Liu2023,loehr2016,dasbiswas2018, Yoshida2023-uw, Kim2023-he,diehl2011}.

In the equilibrium case, the protection phenomenon can manifest itself in the formation of a quasi-particle-like structure in similarity to polaron \cite{Pekar1946,*Landau1948, Frohlich1954, Holstein1959-xs} or solvated ion \cite{Frank1957-xi, Burgess1999-hm} dressed in a polarization coat of the surrounding media. Another example of such a spatial structure can be a complex of a colloidal particle and a nonlinear ``deformation coat'' of liquid crystal formed around it \cite{Lev2002,Foffano2014-et,Musevic2006-os,Ohzono2017-wo}. One of the features of such a quasi-particle formation is the strong coupling or correlation occurring between the particle and its nearest surrounding, such that polarization or deformation coat might protect the inner state of this complex from external fluctuations.

We are interested in  a quasi-particle-like structure in the nonequilibrium case, particularly, in the formation of a similar structure induced by the gas flow scattering on an obstacle.

Steady nonequilibrium structures caused by the scattering of a forced flow on obstacles or impurities represent a broad class of NESSs of many-body systems \cite{liu2020,sonner2017,novak2018,kundu2015,khlebnikov2010,maass2023,schadschneider2010}. The behavior of such systems admits nonequilibrium phase transitions when the spatial NESS structure undergoes a dramatic change at certain critical parameters such as flow velocity, obstacle characteristics, mean gas density, etc. \cite{liu2020,sonner2017,novak2018,kundu2015,khlebnikov2010,maass2023,schadschneider2010}.
The properties of these structures are significantly dependent on whether linear or nonlinear mechanism of scattering is dominant. The same applies to flow-induced forces acting on obstacles and flow-induced nonequilibrium interaction between them, which is usually non-reciprocal, like wake-mediated or nonequilibrium depletion forces \cite{dzubiella2003,wulfert2017,pre2017,ivlev2015}. In the case of a gas, the nonlinear mechanism of scattering can be provoked by the short-range repulsion between particles leading to the blockade effect at certain conditions \cite{kliushnychenko_effect_2014}.

We focus on nonequilibrium phase transitions leading to the formation of nonequilibrium nonlinear structures (like ``strata'' \cite{saitou2012,jaiswal2018,jaiswal2016,bandyopadhyay2022,tsytovich2013,janowski1992,janowsky1994,costin2012}) provoked by the blockade effect in a gas. In the linear case when one can neglect interparticle interaction (low gas concentration, small gas flow,  high penetration of obstacle), the gas density perturbation (wake shape) formed by flow scattering on the obstacle differs slightly from the equilibrium density and can be described in terms of the linear response leading to the asymmetric Green function. In the particular case of advection-diffusion systems, the Green function is given by the Yukawa-like form with asymmetric behavior of screening length that can be interpreted as dynamic screening length\footnote{The term \textit{dynamic screening} is borrowed from the realm of collective energy losses by swift ions in condensed matter \cite{ritchie1982,ritchie1976,pines1956}.} determined by flow velocity \cite{pines1952,thompson1960,rostoker1961}. The asymptotic behavior of the wake profile exhibits exponential growth of gas density ahead of the obstacle and power-law decay of rarefied gas tail behind it \cite{pre2017}, at least in 2D and 3D.

When the interparticle interaction is significant, due to locally increasing gas density, especially ahead of the obstacle, the nonlinear dynamic screening regime becomes dominant, manifesting itself as the blockade effect for a gas with short-range repulsion. This leads to the qualitative change of wake shape or gas density profile, which takes a kink-like form ahead of the obstacle. This corresponds to the formation of a dense gas phase with a sharp boundary. For example, it can be the formation of strata or bow-shocks near a void in dusty plasmas \cite{saitou2012,jaiswal2018,jaiswal2016,bandyopadhyay2022,tsytovich2013} or the formation of a kink-like density profile describing a dense-gas phase ahead of obstacles embedded into the driven hard-core interacting gas or into the driven lattice gas \cite{leung_novel_1994,hipolito_2003,pre2017,pre2018}.

Qualitatively, the growth of a dense gas phase with a sharp boundary (a stratum) can be viewed as the growth of a ``nucleus'' adjacent to the obstacle that serves as a nucleation center. One can suppose that the nucleus surface (domain wall between depleted and dense gas phases), being an additional scatterer for a gas flow, has to protect the state of the nucleus--obstacle complex from the external noise and fluctuations in the rest of the gas.

In this paper, we demonstrate that the nonequilibrium transition to a two-domain gas structure, with a dense phase ahead of the obstacle, is accompanied by the emergence of local invariants which behave like local first integrals.
These invariants describe  the conservation of the gas particle number inside the obstacle and the collective state of the obstacle's nearest gas surrounding.
As a result, this transition leads to the spatial decomposition of the system and to the formation of a quasi-particle-like structure (obstacle complex) caused by the gas flow. We show that the dense gas phase protects the state of the obstacle complex from the noise of external driving field. In addition, noise-induced gas fluctuations are localized  near the domain wall.
We associate this effect with possible nonequilibrium protection effect \cite{kawabata2023,garbe2023,minoguchi2023,cao2021,hu2022,hu2022,shankar2022,bergholtz2021,gong2018}.

To this end, we resort to a simple, though specific, case of one-dimensional driven lattice gas with ring topology. We consider gas scattering on a single partially penetrable impurity, supposing that a driving field that generates the gas flow is non-conservative. We refer to the mean-field Smoluchowsky equations \cite{schutzPTCP2001,chavanis2019,schmittmann1995} for the mean gas occupation numbers  of lattice sites.
This mean-field model is related to the so-called ASEP (asymmetric simple exclusion process) \cite{Galanti2016,maass2023,gouyet2003,mallick2015}. In our approximation, we  neglect fast processes, short-range correlations \cite{tahir-kheli_correlated_1983} in a gas and possible blockade effect due to local fluctuations in a gas \cite{schutzPTCP2001}.

Our paper is organized as follows.
In Sec.~\ref{sec:model}, we introduce the mean-field ASEP model on a ring with a partially penetrable obstacle. In contrast to the well-known Lebowitz-Janowsky blockage model \cite{mallick2015,janowsky1994,costin2012,janowski1992}, an obstacle is given by an impurity site with reduced possible value of vacancies for gas particles.
In Sec.~\ref{NET}, we describe the nonequilibrium phase transition between two classes of spatial NESS structures. The first class describes a nonordered gas phase characterized by slight accumulation of  gas particles near obstacle's front side. The second one describes an ordered phase corresponding to two-domain gas structures with the kink form of density profile.
The transition into two-domain structures  occurs when the values of mean gas concentration, driving field, and impurity penetration (or capacity) exceed  certain critical values. We show that this transition is accompanied by  the spatial decomposition of the system. The occupation of the impurity site and the total occupation of its edges exhibit invariant behavior becoming insensitive to variation of main system parameters such as driving field, mean gas concentration, the number of particles, ring size.
In Sec.~\ref{sec:III-LIOMs}, we consider a particular case of a time-fluctuating driving field to demonstrate that the emerged invariants behave like the local first integrals and/or adiabatic invariants. We also show that gas density fluctuations induced by the external field noise are suppressed near the impurity and are strongly localized near the interface between the depleted and dense gas phases, Sec.~\ref{IIIB}. This spatial decomposition of the system  points at the possible nonequilibrium protection effect. The transitions between different NESSs at the sudden change of the driving field are considered in Sec.~\ref{sec:gamma}. We show that the relaxation rates from one NESS to another are different, except for the states with the opposite direction of the driving field, and they take real and complex values in the over- and subcritical regions, respectively. Such an inter-state transitions are realized by one-step or multi-step scenarios implying the sequential generation of shock waves in time that leads to a step-by-step reconfiguration of the system. Concluding remarks are given in Sec.~\ref{sec:conclusion}.

\section{\label{sec:II-Ring}Nonequilibrium transition to a two-domain gas structure on a ring }
\subsection{\label{sec:model}Mean-field ASEP model on a ring with partially penetrable obstacle. Local-equilibrium coarse-graining}

Our consideration is based on the approach known as the local equilibrium approximation.
This approach is often exploited to describe kinetics of fluctuations on macroscopic time scale neglecting fast processes, in particular, to describe fluctuations in electron-phonon systems, electron-hole plasma,  fluctuations in the system of adsorbed atoms in the  potential relief of a solid-state substrate, e.g., \cite{Chumak1980,chumak1992book,Chumak_Uebing,richards1977,zubarev1973,Ropke2013}.

The idea of this approach is in the existence of two different time scales (the fast and slow times) of system relaxation to its equilibrium state. During the fast time, the local spatial equilibrium establishes. Further, the relaxation is governed by slow processes and goes on macroscopic scales.   

We apply this approach in combination with the mean-field approximation to describe long-time relaxation kinetics near NESS. We resort to the limiting case of two-component driven lattice gas,
that is phenomenological model of adsorbed atoms in the potential relief of solid-state substrate. In the end of this part, we consider the particular case of a narrow channel that can be reduced to the quasi-1D lattice case.  Details about  the approach and used approximations are given in Appendix~\ref{sec:appendix_3}.

One gas component represents mobile particles jumping between nearest lattice sites $k$ and $j$, with  mean jump frequencies, generally,  $\nu_{kj}\neq \nu_{jk}$.
The other (heavy) component is the static one describing distribution of impurities, assuming that each lattice site can be occupied by only one gas particle.
The state of the system is described in terms of the site occupation numbers  $\hat{n}_k=\{0,1\}$ of mobile particles, and $\hat{u}_k=\{0,1\}$ of impurities, $\hat{n}_k+\hat{u}_k=\{0,1\}$, $k$ is site index.
In our case, the local equilibrium approximation means that during some time interval $\tau'$ (the long time scale), satisfying the condition $\tau'\langle \nu_{kj}\hat n_k(1-\hat u_j - \hat n_j)\rangle\gg 1$, see \cite{Chumak1980,chumak1992book} and Appendix~\ref{sec:appendix_3}2, the equilibrium of each site with its nearest environment establishes, i.e., many particles visit the site.
Such coarse-graining allows us to correctly introduce the fluctuations $\delta n_k (t)$ as a small value for long-time scales $t\approx \tau'$.\footnote{For discrete $\hat{n}_k=\{0,1\}$, the values $\delta \hat n_k=\hat n_k-\langle \hat n_k\rangle$ are not small in general, since the root-mean-square fluctuation of the concentration of any site is $\left\langle \delta \hat n_k^2 \right\rangle = \left\langle \hat n_k^2 \right\rangle - \left\langle \hat n_k \right\rangle^2 = \left\langle \hat n_k \right\rangle - \left\langle \hat n_k \right\rangle^2 $, as a result, the relative magnitude $\left\langle \delta \hat n_k^2 \right\rangle / \left\langle \hat n_k \right\rangle^2\sim1/ \left\langle \hat n_k \right\rangle$ can be quite large.}
We also avoid difficulties with the introducing of the time derivative that is problematic when describing the evolution of discrete variable $\hat{n}_k$, see \cite{tahir-kheli_correlated_1983,richards1977,shumilin2019}.

One of the methods to eliminate fast processes is the method of the local equilibrium statistical operator \cite{zubarev1973,Ropke2013}, when the system is averaged by the statistical operator whose form is similar to the one of equilibrium statistical operator, but with the local chemical potential that depends on the slow-time variable.
This approach enables one to describe the kinetics of the fluctuations $\delta n_k(t)$ by Langevin equation near equilibrium state $n^\text{e}_k$ of the system, see \cite{Chumak1980,chumak1992book,Chumak_Uebing}.

However, we are interested in the nonequilibrium steady states and fluctuations near these states. To this end, we apply the mean field approximation in the framework of the local equilibrium approach, see Appendix~\ref{sec:appendix_3}.
This gives us a the possibility  to describe the slow-time dynamics of the mean occupation numbers $0\leq n_k(t)\leq1$, $n_k(t)={\langle\hat n_k\rangle}_t\equiv\langle\hat n_k\rangle$, in the  form that is similar to the one of the mean field Smoluchowski equation, see \cite{chavanis2019,schmittmann1995,leung_novel_1994,pre2017}, 
\begin{equation}\label{eq:8intext}
  \partial_t  n_k = \sum_j ( \nu_{jk} n_j h_k - \nu_{kj}n_kh_j ),
\end{equation}
where $h_k=1-u_k-n_k$, and $u_k=\hat{u}_k=\{0,1\}$ is given distribution of impurity particles, $\nu_{kj}$ is the mean frequency of jumps of the mobile gas particles from $k$th site to its neighboring $j$th one.
The asymmetry in forward-backward jumps $\nu_{kj}\neq\nu_{jk}$ is caused by external drive, that will be specified below.

We can also  describe the fluctuations $\delta n_k$ near nonequilibrium steady states $n^s_k$  of the system, given by Eq.~(\ref{eq:8intext}), by the Langevin equation that, for  small $\delta n_k$, takes the form, see Appendix~\ref{sec:appendix_3},
\begin{align}
 \partial_t \delta  n_k=&
 \sum_j \left[\nu_{jk} \left(h_k^s\delta n_j -n_j^s\delta  n_k\right)
    -\nu_{kj}\left( h_j^s\delta n_k-n_k^s\delta  n_j\right)\right]\nonumber\\
    &+ \delta \tilde I_k,
\end{align}
with correlation function of the Langevin source that, for rare jumps, can be written as
\begin{multline}\label{eq:LangevinCor_2}
\left\langle \delta \tilde I_k(t) \delta \tilde I_{k'}\left(t'\right)\right\rangle \approx
\delta\left(t-t'\right) \\
\times\sum_j \left(\nu_{kj} n^s_k h^s_j + \nu_{jk} n^s_j h^s_k\right)\left(\delta_{kk'}-\delta_{jk'}\right).
\end{multline}
Obtained equations describe the relaxation kinetics of the non-equilibrium 2D- or 3D-system only for long-time scales, when any fast processes and short range correlations can be considered as insignificant.
Note that we can get a more general form of the equation describing the system kinetics (see Appendix~\ref{sec:appendix_3}) that is similar to the stochastic mean field Smoluchowski equation \cite{chavanis2019,schmittmann1995,leung_novel_1994,pre2017}.

In what follows, we consider the particular case of a lattice in the form of a narrow channel with longitudinal $L_{\parallel}$ and transverse $L_{\perp}$ sizes, so that $L_{\parallel}\gg L_{\perp}$. We also assume that external driving field is applied along the channel.
Considering the channel as a set of transverse cells located along its longitudinal direction we can reduce the system to the quasi-1D lattice case by the averaging of Eq.~(\ref{eq:8intext}) over transverse coordinate, see Appendix~\ref{sec:appendix_3}.
As a result, the mean-field dynamics is governed by the equations of the same form as Eqs.~(\ref{eq:8intext}), where the number of 1D-lattice site $k$ corresponds to transverse cell number, the mean occupation number $n_k$ corresponds to the the mean particle concentration in the $k$th cell, $U_k$ describes obstacle corresponding to the mean concentration of impurities in the cell, $h_i=1-n_k-U_k$ is concentration of vacancies, see Eq.~(\ref{eq:n_h_U_A}) in Appendix~\ref{sec:appendix_3}, and $j=k\pm1$.
\begin{figure*}[t]
    \includegraphics[width=.95\textwidth]{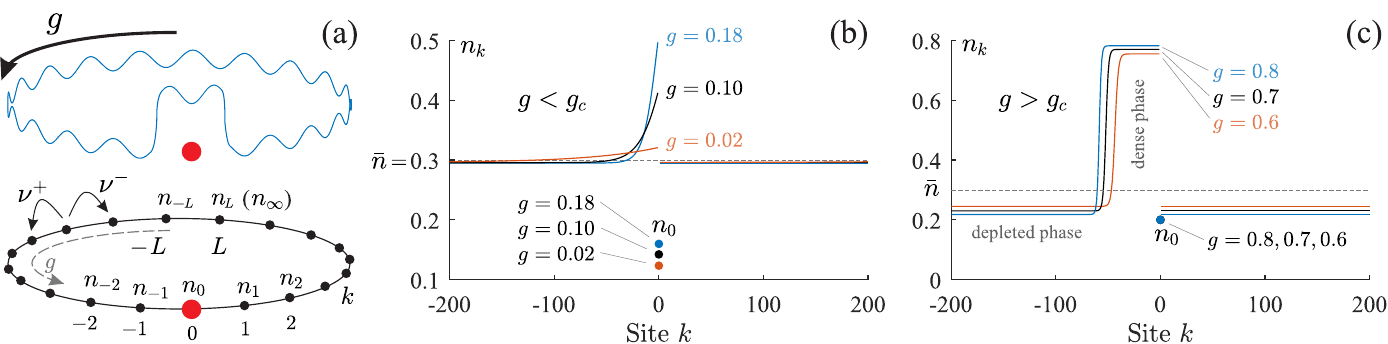}
    \caption{\label{fig:1}(a) Periodic chain with ring topology and the impurity site at the origin (below): $\nu^\pm=\nu(1\pm g)$ are particle hopping rates along and against the direction of nonconservative field $g$.
    Qualitatively, this ``hopping transport'' model can be associated with a potential energy landscape as illustrated (above).
    (b) Typical NESSs in subcritical regime, at $\bar{n}=0.3$, $U=0.6$, and $g=0.02,0.1,0.18<g_c$.
    (c) Typical NESSs in overcritical regime at $\bar{n}=0.3$, $U=0.6$, and $g=0.6,0.7,0.8>g_c$. The number of lattice sites $L_0=401$, the ring length $2L=400\ell$, with lattice constant $\ell$.
    The distributions in (b) and (c) were obtained from direct numerical solutions of the mean-field Eq.~(\ref{eq:nk})--(\ref{eq:Jk,k+1}) as steady-state profiles established after $\approx1.6\times10^7$ time steps of evolution since the driving field was switched from $g_0=0$ to $g$ at $t_0=0$. The resulting NESS was regarded as finally established if $\max_k[n_k(\tau)-n_k(\tau-\Delta\tau)]\leq10^{-30}$ with $\Delta\tau=0.01$, where $\tau=\nu t$, see Appendix~\ref{sec:appendix_3}6 on details of numerical implementation.}
\end{figure*}
To demonstrate features of nonequilibrium transition to blockade regime in a gas caused by flow scattering on impurities we resort to the specific 1D-lattice case  with ring topology and penetrable obstacles.
We  introduce the obstacle directly as the cell doped by impurity (heavy) atoms. In the quasi-1D limit this cell corresponds to the impurity site that is characterized by the mean concentration $U$ of impurity atoms in the cell.
Since $U<1$ one can regard the impurity site as the partially penetrable one.

In this setting, the gas kinetics is given by equations for mean occupation numbers $n_k(t)$:
\begin{equation}\label{eq:nk}  
  \dot{n}_k=J_{k-1,k}-J_{k,k+1},
\end{equation}
where $J_{k,k+1}$ is bond current between $k$ and $k+1$ sites:
\begin{equation}\label{eq:Jk,k+1}
J_{k,k+1}=\nu^+ n_k(1-n_{k+1}-U_{k+1})-\nu^- n_{k+1}(1-n_{k}-U_{k}),
\end{equation}
where $\nu^{\pm}=\nu \pm \delta \nu$ are forward-backward particle hopping rates between nearest sites. The asymmetry $\pm \delta \nu=\pm \nu g$ is caused by the action of nonconservative driving force or field $g$. The nonconservative force is usually inherent for transport problems on a ring, as it is for colloidal particle in a periodic potential, see  \cite{Seifert2012}.
In the case of an infinite system, driving field can be given by conservative force that leads to asymmetric particle jumps $\nu_{ji}\approx\nu[1+\bg\cdot(\br_i-\br_j)/\ell]$ where  $|\bg|=\ell|\bG|/(2kT)<1$, $\ell$ is the lattice constant, and $\bG$ is external drive, see \cite{pre2017}. This form corresponds to the one of the wind force often encountered in  electromigration  of adsorbed atoms on a solid-state substrate, see, e.g., \cite{krug1997,ishida1994,Verbruggen1988}.

In contrast to most ASEP-based models of the blockade effect (including numerous versions \cite{janowski1992,janowsky1994,costin2012,kolomeisky_1998,GREULICH20081972,Greulich_2008,Soh2018,Soh2017,cirillo2016,sarkar2014,lobaskin2024}), where obstacles often realized via defect bonds, e.g., slow bonds,
we implement the obstacle as the partially transparent impurity-site\footnote{Sometimes, the terms like ``defective sites'' are  associated with the set of distinct bonds with modified hopping rates. To avoid confusion in terminology, we use the term ``impurity site'' instead of the  used terms ``slow site'' or ``defective site'', see \cite{Greulich_2008,janowski1992}.} that corresponds to the channel cell partially occupied by impurity (heavy) gas particles with concentration $U$.
Such   obstacle manifests itself in decreasing of occupation probability of this site by a gas particle (due to the decreasing of possible vacancies $1-U$), and in the reduction of transition rates to this site from nearest neighbor sites.
Qualitatively, $U$ can be associated with  effective potential created by impurity atoms in the cell corresponding to the impurity site, as illustrated in Fig.~\ref{fig:1}(a).
This potential is entailed by strong short-range (hard-core) repulsion between adsorbed atoms  in the same minimum of potential relief.
Another difference from standard ASEP-like blockage problem setups is in the use of so-called exclusion process with external field \cite{Bertini2015-mf,espigares2013,sellitto2008}. In this case, the transition to a blockade regime occurs not only at certain critical values of mean gas concentration and obstacle transparency parameter, but also critically depends on the driving field value.

Note again that under used approximations  we neglect fast processes, short-range correlations, local memory effect \cite{tahir-kheli_correlated_1983}, and possible nonequilibrium transition induced by local fluctuations in a gas \cite{bodineau2005,espigares2013,hurtado2011}.

For simplicity, we consider a particular case of single inclusion and neglect  effects entailed by collective scattering. The latter would lead to a number of specific features caused by the additional induced correlations between impurities \cite{winpr}. The impurity is placed  at zeroth lattice site $U_k=U\delta_{k,0}$, and the  numbering of lattice sites is taken as $-L\leq k \leq L$, see Fig.~\ref{fig:1}(a). The total number of ring sites is $L_0=2L+1$. Equations~(\ref{eq:nk}) are supplemented by periodic boundary conditions $n_{L+1}=n_{-L}$.

\subsection{\label{NET}Nonequilibrium transition to two-domain gas structure}
We consider the nonequilibrium transition emerging between the main two types (or classes)  of nonequilibrium, steady-state, spatial gas structures which can be formed by flow scattering on the obstacle at $t\rightarrow \infty$.
These NESSs are determined by the system's parameters: the driving field $g$, the mean gas concentration $\bar{n}=N/L_0$ ($N$ is the particle number), and impurity capacity $1-U$. We show that the transition to the two-domain gas structure with dense and depleted phases is accompanied by the emergence of quantities which become insensitive to variations of the main parameters driving the system, i.e., behave like invariants. We also estimate critical values $(\bar{n}_c, g_c, U_c)$ and changes in the asymptotic behavior of order parameters, which characterize this transition.

In the steady-state case, $\dot n_k=J_{k-1,k}-J_{k,k+1}=0$, the system has an additional conserved quantity, besides the particle number $N$ (or mean gas concentration $\bar{n}$). It is a current $J_{k,k+1}=J=\mathrm{const}$, Eq.~(\ref{eq:Jk,k+1}), flowing through  each  bond $(k,k+1)$ of the ring
\begin{equation}\label{eq:4}
\nu^+ n_k(1-n_{k+1}-U\delta_{k+1,0})-\nu^- n_{k+1}(1-n_{k}-U\delta_{k,0})= J
\end{equation}
written for the case of single impurity site at $k=0$, i.e., $U_k=U\delta_{k,0}$.

One can qualitatively estimate the value of $J$. For a sufficiently large ring $L\gg1$, one can suppose that far from the impurity, $|k|\gg1$, the site-occupancies are approximately equal, i.e., $n_{-L}\approx n_L\approx n_\infty$, that gives
\begin{equation}\label{eq:J infty}
  J_\infty= J_{L,-L}\approx (\nu^+-\nu^-)n_\infty(1-n_\infty).
\end{equation}

There are  two main classes of nonequilibrium steady structures formed in the system at $t\rightarrow \infty$. The first one is resulted by flow scattering on obstacle at relatively small values of $\bar{n}$, $g$, and $U$, see Fig.~\ref{fig:1}(b). The second class is described by kink-type spatial structures of NESS, and corresponds to emergence of two-domain gas structure when parameters $(\bar{n},g,U)$ exceed certain critical values $(\bar{n}_c,g_c,U_c)$, see Fig.~\ref{fig:1}(c). Note that such a structure is the effect of the ring topology. In contrast, in the case of an infinite chain, the gas density perturbation caused by flow scattering can be located around the obstacle and dumps to equilibrium state at infinity.

Such behavior of gas concentration along the chain can be qualitatively explained.
In the subcritical case ($g<g_c$), the external field $g$ winds gas particles towards the end of chain corresponding to the one of the obstacle edges, that  is analogical to skin effect with a slight accumulation of the particles near the boundary of a chain \cite{shankar2022}. Far from obstacle, the gas concentration have to weakly differ from its equilibrium value $n_\infty \approx n^\text{e}\approx \bar n$, as a result, the steady state current $J_\infty$ weakly differs from current for obstacle-free gas, similarly as in \cite{lobaskin2024,lobaskin2022}. In other words, the obstacle effect on the system is spatially  localized.

In the case of overcritical regime ($g>g_c$) characterized by two-domain gas structure, we can say about some saturation effect with respect to the field $g$.

For finite $g$ ($g>g_c$) the profile of two-domain structure (domain sizes, their densities) depends on $g$, except for the the impurity site $k=0$, whose occupation $n_0(g)=\text{const}$ is insensitive to the field $g$, Fig.~\ref{fig:1}(c).
In this context, $g_c$ plays the role of saturation field for impurity-site state.
To explain the saturation of impurity site $k=0$ by gas particles
we resort to the ``spin'' representation $\sigma_k$ via the replacement $n_k=(1-U_k)/2+\sigma_k$ (the formal correspondence Lattice-Gas---Ising models \cite{baxter1982}). The dense gas phase is restricted by two domain walls.
The first wall (kink) is located far from inclusion site, $k=0$, and position of its center $k_s$ is determined by the applied field $g$.
The center of other wall (anti-kink) is pinned by the impurity site $k=0$.
Intuitively, the positive spins ($\sigma_k>0$) describe the dense gas phase, and negative ones ($\sigma_k<0$) do the rarefied gas phase. The value $\sigma_{k_*}=0$ has to correspond to the center of a domain wall separating two phases, and located at site $k_*$. It means that the gas concentration at the centers of domain walls corresponds to the half-filling of the available vacancies for these sites $k_*$.
As a result, $n_{k_s}=1/2$ for the wall located far from inclusion ($U_{k_s}=0$), and  $n_0=(1-U)/2$ for the wall pinned by impurity.
Below we show that this is indeed the case.

\medskip
In what follows, we restrict our consideration mainly to the  case $\bar n<1/2$. Analogous description in $\bar n >1/2$ regime is straightforward by virtue of wake inversion effect \cite{pre2017,kliushnychenko_effect_2014}, stemming from particle-hole symmetry, when the kink profile describes a cavity behind the obstacle. This case corresponds to what is called high-density (HD) phase in ASEP models \cite{mallick2015}. In particular, the symmetry relative to $\bar n \leftrightarrow 1-\bar{n}$ interchange is evident from numerical calculations in the $\bar{n}\in[0\ 1]$ range, as is shown in this subsection later.

\medskip
\textit{Subcritical region.} In the equilibrium case (in the absence of driving field, $g=0$), the current $J=0$. On the average, the gas has no structure,  akin to quasi-one-dimensional liquid in equilibrium \cite{ziman1979}. The equilibrium distribution of  mean occupation numbers $n_k$  is given by expression
\begin{eqnarray}\label{eq:n49fj28HsI2h}
  n_k^\text{e} &=& (1-U)n^\text{e}\delta_{k,0} + n^\text{e}(1-\delta_{k,0}),
\end{eqnarray}
where $n^\text{e}=(1-\overline{U})^{-1}\bar{n}$, and $\overline{U}=U/L_0$. At small driving field $g\ll1$, and/or small mean gas concentration $\bar{n}\ll1$, and/or high impurity penetrability $U\ll1$, the distribution of gas density perturbation $\delta n_k$, caused by flow scattering on the obstacle, weakly deviates from its equilibrium value $n^\text{e}$ (or mean gas concentration $\bar{n}$) for a large ring, $L_0\gg1$. The scattering leads to the relatively small accumulation of gas particles ahead of the obstacle (the impurity site), see Fig.~\ref{fig:1}(b).

This class of steady-state structures or solutions can be easily estimated by resorting to the continuum limit of Eqs.~(\ref{eq:nk}) for gas density $n(x)$ outside impurity site, see Appendix~\ref{sec:appendix}. In the lowest order of the long-wave approximation, $n(x)$ is governed by the Burgers equation.

Representing gas concentration $n(x)=\bar{n}+\delta n(x)$ outside the inclusion and occupancy of impurity site  $n_0=\bar{n}+\delta n_0$ as small deviations from mean gas concentration $\bar{n}$ one can obtain linearized equations corresponding to Eq.~(\ref{eq:4})
\begin{subequations}\label{eq:J_linear}
  \begin{equation}\label{eq:J_linear a}
    -\partial_x \delta n + 2g(1-2\bar{n})\delta n=\delta J,
  \end{equation}
  \begin{gather}\label{eq:J_linear b}
    \delta n_{-1}\left[\nu^+(1-U-\bar{n})+\nu^-\bar{n}\right]-
    \delta n_0\left[\nu^+\bar{n}+\nu^-(1-\bar{n})\right]\nonumber\\
    =\delta J+\nu^+U\bar{n},
  \end{gather}
  \begin{gather}\label{eq:J_linear c}
   \delta n_0\left[\nu^+(1-\bar{n})+\nu^-\bar{n}\right]-\delta n_{1}\left[\nu^+\bar{n}+\nu^-(1-U-\bar{n})\right]\nonumber\\=\delta J-\nu^-U\bar{n}.
  \end{gather}
\end{subequations}
Here we change coordinate system for dimensionless variable $-2L<x<0$ so that occupancies on impurity edges are written as $\delta n_{-1}=\delta n(0-)$ and $\delta n_{1}=\delta n(-2L+0)$. These equations are supplemented by one for conservation of the particle number (or $\bar{n}$) to obtain unknown constants $\delta n_{-1}$, $\delta n_{1}$, $\delta n_{0}$ and $\delta J=J-2g\bar{n}(1-\bar{n})$.

\begin{figure}[t]
    \includegraphics[width=.85\columnwidth]{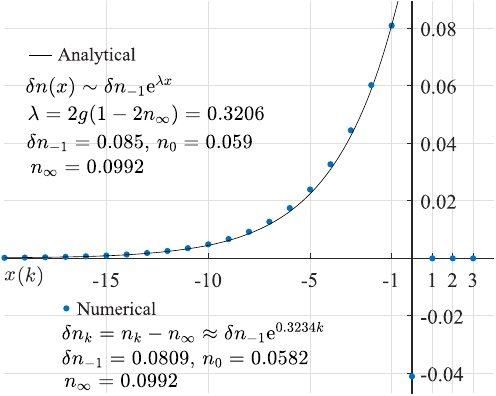}
    \caption{\label{fig:2a*}
    Behavior of density distribution (NESS profile) ahead of the obstacle in subcritical regime at $\bar{n}=0.1$, $U=0.6$, and $g=0.2<g_c$: comparison of numerical result and analytical asymptotics, Eq.~(\ref{eq:del n x}). Here $x$ corresponds to continuous coordinate, $k$ is the site index, and $L_0=401$. The distribution of $n_k$ was obtained from the numerical solution of mean-field equations~(\ref{eq:nk})--(\ref{eq:Jk,k+1}) in the same way as in Fig.~\ref{fig:1}, see also Appendix~\ref{sec:appendix_3}6 for details.}
\end{figure}

\begin{figure}[t]
    \includegraphics[width=.9\columnwidth]{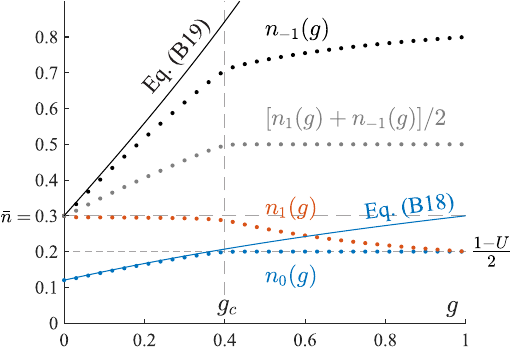}
    \caption{\label{fig:3a*}
    Occupations of impurity site $n_0$, its edges $n_{\pm1}$, and their half-sum as a function of external field $g$. Above critical field $g_c$, determining the phase transition to kink-profile NESS, quantities $n_0$ and $n_{-1}+n_1$ are no longer dependent on $g$. Here, $\bar{n}=0.3$, $U = 0.6$, and $L_0=401$.
    The results are obtained based on the numerical solution of mean-field equations~(\ref{eq:nk})--(\ref{eq:Jk,k+1}) in the same way as in Fig.~\ref{fig:1}, see also Appendix~\ref{sec:appendix_3}6 for details.}
\end{figure}

In zeroth order in $\lambda/L_0 \ll 1$, see Appendix~\ref{sec:appendix}, behavior of $\delta n(x)$ takes the simplest form
\begin{eqnarray}\label{eq:del n x}
  \delta n\approx \delta n_{-1}\exp(\lambda x),
\end{eqnarray}
where $\lambda^{-1}=2g(1-2\bar{n})$ is dynamic screening length. This approximation neglects the difference between $n_\infty$ and $\bar{n}$  far away from the obstacle (actually $n_\infty<\bar{n}$ ). As a result, $\delta J\approx \delta n_1 \approx 0$.
Nevertheless, obtained results for $\delta n(x)$ as well as  for $\delta n_0(\bar{n},g,U)$ and $\delta n_{-1}(\bar{n},g,U)$ [expressions for which are given by Eqs.~ (\ref{eq:A:delta n 0 zer}) and (\ref{eq:A:delta n-1 zer})] are in more-or-less good agreement with asymptotic numerical results obtained from Eqs.~(\ref{eq:nk}) at long times, see Figs.~\ref{fig:2a*} and \ref{fig:3a*}.

In subcritical domain, the class of nonequilibrium steady structures of the gas or the class of stationary solution of the system Eqs.~(\ref{eq:nk}) correspond to the linear dynamic screening regime.
The spatial structure of the gas exhibits exponential growth region, due to particle accumulation ahead of the obstacle, with characteristic length $\lambda^{-1}$, while behind the obstacle, we have small density deviation from the equilibrium value $n^\text{e}\approx \bar{n}$. 
Such particle accumulation near the obstacle edge is the simplest manifestation of non-Hermitian skin effect, see \cite{shankar2022,Liu2023}.

\textit{Overcritical region.} If the system parameters $\bar{n}$, $U$, and $g$ exceed certain critical values  $(\bar{n}_c, U_c, g_c)$, the system undergoes transition to two-domain structure with depleted and dense gas phases Fig.~\ref{fig:1}(c).
The spatial structures of NESS behave similarly to a kink solution.

To find critical estimates of $(\bar{n}_c, U_c, g_c)$ and to assess the behavior of the main characteristics of this transition, we resort to the rough approximation of a solution\footnote{
In this case, the long-wavelength approximation (usual Burgers equation) is not sufficient since the width of the domain wall pinned by impurity site is of the order of the lattice constant $\ell$.}
representing it as a kink with sharp step
\begin{equation}\label{eq:step}
n_k=\left\{\begin{array}{l}
n_{\infty},\quad\quad\ \ -L\leq k< -S\\
1-n_\infty,\quad-S\leq k<0\\
n_0,\quad\quad\quad  k=0 \\
n_{\infty},\quad\quad\ \  0<k\leq L.
\end{array}\right.
\end{equation}
In the case of a single impurity, the class of the approximate kink-type solutions, Eq.~(\ref{eq:step}), is determined by concentration $n_\infty$, the mean occupation number of the impurity site $n_0$ and the length of dense phase (stratum) $S$ corresponding to the lattice site $k=k_s=-S$.

Note that at $k_s=-S=-L$ and $n_0=(1-U)/2$, Eq.~(\ref{eq:step}) gives exact particular single-kink solution for a finite chain. Generally, result is  sensitive to the choice of boundary condition for a finite unclosed chain.

Under approximation Eq.~(\ref{eq:step}), $n_\infty$ and $n_0$ can be obtained from equations $J_{-1,0}=J_\infty$ and $J_{0,1}=J_\infty$, see Eq.~(\ref{eq:4}),
\begin{subequations}\label{eq:J-1,0 J0,1}
  \begin{equation}\label{eq:J-1,0 J0,1 a}
    \nu^+ n_{-1}(1-n_0-U) - \nu^-n_0(1-n_{-1}) = J_\infty,
  \end{equation}
  \begin{equation}\label{eq:J-1,0 J0,1 b}
    \nu^+ n_0(1-n_1) - \nu^-n_1(1-n_0-U) = J_\infty,
  \end{equation}
\end{subequations}
where $n_{-1} = 1-n_\infty$ and $n_1 = n_\infty$ are the occupation numbers of the nearest sites from the left and to the right of the impurity, correspondingly, and the current $J_\infty$ is given by Eq.~(\ref{eq:J infty}). Equations (\ref{eq:J-1,0 J0,1}) can be rewritten in the form
\begin{subequations}\label{eq:n-infty-n0}
  \begin{equation}\label{eq:n-infty-n0 a}
    \left\{ \nu^+(1-n_\infty) + \nu^-n_\infty \right\} (1-U-2n_0) = 0,
  \end{equation}
  \begin{eqnarray}\label{eq:n-infty-n0 b}
    \left\{\nu^+(1-n_\infty)-\nu^-n_\infty\right\}(1-U) \nonumber\\
    = 2(\nu^+-\nu^-)n_\infty(1-n_\infty).
  \end{eqnarray}
\end{subequations}
It follows directly from Eq.~(\ref{eq:n-infty-n0 a}) that the mean occupancy of impurity site by gas particles
\begin{equation}\label{eq:no(U)}
  n_0 = \frac{1-U}{2}
\end{equation}
is determined only by the impurity capacity $(1-U)$. It does not depend on the number or mean concentration $\bar{n}$ of particles, on the value of driving field $g$ and as a result on the current $J_\infty$ flowing through the obstacle, in contrast to the subcritical regime, Figs.~\ref{fig:3a*} and \ref{fig:4b}.
\begin{figure}
    \includegraphics[width=\columnwidth]{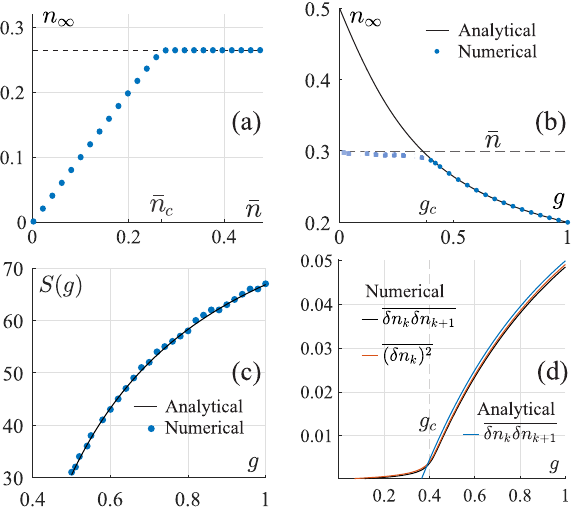}
    \caption{\label{fig:abcd}
    (a) Plot shows the existence of critical concentration value $\bar n_c$ above which $n_\infty$ no longer depends on $\bar n$, signifying the transition into two-domain structure at $\nb>n_c$ (for $g=0.5$).
    (b) Dependence of concentration $n_\infty$ on field $g$. Analytical solution for $n_\infty$ is given by Eq.~(\ref{eq:jjfkdjwi}).
    (c) Dependence of domain wall position $k_s=-S(g)$ or the spatial width of dense phase $S(g)$ on the driving field $g$ (numerical against Eq.~(\ref{eq:S})). The numerical values of $S$ were taken as the closest site to the domain wall $k_s$ defined by $n(k_s)=1/2$, cf.~\cite{pre2018} for domain wall stopping criterion.
    (d) Inter-particle correlations at the nearest sites $\overline{\delta n_k \delta n_{k+1}}$, where $\delta n_k = n_k - \bar n$, and $\overline{(\dots)} = L_0^{-1}\sum_k(\dots)$; analytical curve is given by Eq.~(\ref{eq:corre}).
    Gas concentration $\bar n=0.3$ for (a), (c), and (d);  $U=0.6$ and $L_0=401$ for all panels.
    The results are obtained based on the numerical solution of mean-field equations~(\ref{eq:nk})--(\ref{eq:Jk,k+1}) in the same way as in Fig.~\ref{fig:1}, see also Appendix~\ref{sec:appendix_3}6 for details.}
\end{figure}
\begin{figure*}[t]
    \includegraphics[width=.9\textwidth]{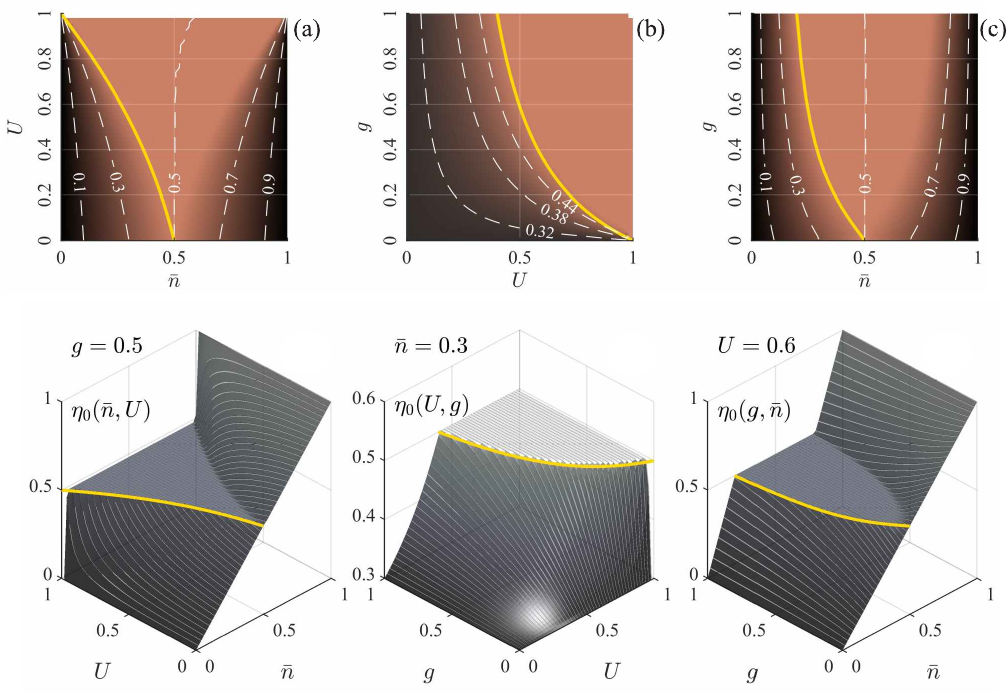}
    \caption{\label{fig:4b} (Upper row) Two-dimensional projections of the critical surface for the phase diagram given in $(U,\nb,g)$ parameter space: (a) $(U,\bar n)$-projection at $g=0.5$,
    (b) $(g,U)$-projection at $\bar n=0.3$, and
    (c) $(g,\bar n)$-projection at $U=0.6$.
    (Lower row) Corresponding behaviors of $\eta_0=n_0/(1-U)$ as a function of the same pairs of parameters. The plateaux at $\eta_0=0.5$ correspond to the kink-phase (two spatial domains).
    Analytically estimated phase boundaries (solid lines) are given by
    (a) $U(\bar n)=1-[4\bar n(1-\bar n)]/(3-4\bar n)$,
    (b) $U(g) = 1-[4gn(1-n)]/(1-2n+g)$, and
    (c) $g(\nb) = (1-U)(1-2\nb)/[4\nb(1-\nb) + U -1]$.
    Ring length is $2L=400\ell$ for all cases. Panels (a) and (c) features inversion symmetry in $\bar{n}$ between high-density (HD) and low-density (LD) phases known in ASEP models \cite{mallick2015}.
    The results are obtained based on the numerical solution of mean-field equations~(\ref{eq:nk})--(\ref{eq:Jk,k+1}) in the same way as in Fig.~\ref{fig:1}, see also Appendix~\ref{sec:appendix_3}6 for details.}
\end{figure*}
The value of $n_0$ saturates at the nonequilibrium transition and corresponds to the half-filling of the impurity site.
Note that half-filling is a property of the center of domain wall separating dense and depleted phases. In addition, the value of $n_0$  also does not depend on  amplitude $1-2n_\infty$ and length $S$ of the kink.

In other words, saturation or half-filling of the impurity $n_0/(1-U)=1/2$ indicates the emergence of ordered two-domain structure in the system and can serve as invariant that characterizes the class of kink-type solutions, despite their form that is determined by three parameters $\bar{n}, U, g$, see~Fig.~\ref{fig:1}(c).

Equation~(\ref{eq:n-infty-n0 b}) has two distinct solutions from which only one,
\begin{equation}\label{eq:jjfkdjwi}
  n_\infty(g) = \frac{1}{2} \left\{ 1 + \frac{n_0}{g} - \left[ 1-2n_0+\left(\frac{n_0}{g}\right)^2\right]^\frac{1}{2} \right\},
\end{equation}
satisfies the condition $0\leq n_\infty\leq 1$,
where $n_0=n_0(U)$ is given by Eq.~(\ref{eq:no(U)}), and $\nu^{\pm}=\nu (1\pm g)$ is taken into account.  As is seen, concentration $n_\infty$,  determining the amplitudes of two gas phases, loses its dependence on the total number of particles $N$ in the system or mean gas density $\bar n$ (for $\bar n>n_c$), Fig.~\ref{fig:abcd}(a). Figure~\ref{fig:abcd}(b) is the comparison of $n_\infty(g)$ estimate against numerical result showing a good agreement.

Using Eq.~(\ref{eq:jjfkdjwi}) one can easily estimate the length $S$ of dense gas phase or position of the second domain wall (the first one coincides with impurity site). Taking into account the particle number conservation and our approximation for $n_k$ Eq.~(\ref{eq:step}) one can write
\begin{equation}\label{eq:j38gyqxkYEnvvY6}
  \sum_{k=-L}^{L} n_k = (2L-S)n_\infty + S(1-n_\infty) + n_0 = N.
\end{equation}
It enables to write expression for fraction of the ring $S/L_0$ occupied by dense phase
\begin{equation}\label{eq:S}
  \frac{S}{L_0} = \frac{\bar n - n_\infty}{1 - 2n_\infty} + \frac{1}{L_0}\frac{n_\infty-n_0}{1-2n_\infty}
  \approx \frac{\bar n - n_\infty}{1 - 2n_\infty}.
\end{equation}

For a large ring, $L_0\gg1$, the part of the ring occupied by dense phase does not depend on its size
and is determined only by mean concentration $\bar{n}$ and concentration at infinity $n_\infty$. This gives a good agreement with numerical data, Fig.~\ref{fig:abcd}(c). The latter is the result of strong correlations between gas particles in the presence of driving field within ring topology --- when external field $g$ gathers gas particles from all over the system towards impurity.

To demonstrate enhancement of inter-particles correlation after the  nonequilibrium transition we consider correlation at the nearest sites
\begin{equation}\label{eq:n94m1cjs5H34}
  \overline{\delta n_k \delta n_{k+1}} = \overline{n_kn_{k+1}} - \bar n^2,
\end{equation}
where $\overline{(\dots)} = L_0^{-1}\sum_k(\dots)$, and $\delta n_k = n_k - \bar n$.
Taking into account
$$\overline{J_{k,k+1}} = J_\infty = (\nu^+-\nu^-)n_\infty(1-n_\infty),$$
it is easy to show that
\begin{eqnarray}\label{eq:prevjg89}
\overline{n_kn_{k+1}} &=& \bar n - n_\infty(1-n_\infty) -\nonumber\\
&-& \frac{1}{\nu^+-\nu^-}
\overline{\left( \nu^+n_kU_{k+1} - \nu^-n_{k+1}U_k \right)}.
\end{eqnarray}
In case of single impurity ($U_k = \delta_{k,0}$), the last term in Eq.~(\ref{eq:prevjg89}) reads $(U/2gL_0)\left[ (1-2n_\infty)+g\right]$.
For a large ring ($L_0\gg1$), this term can be neglected and expression (\ref{eq:n94m1cjs5H34}) takes the form analogous to gas compressibility, see \cite{spohn1991},
\begin{eqnarray}\label{eq:corre}
  \overline{\delta n_k\delta n_{k+1}} &\approx&
   \bar n(1-\bar n) - n_\infty(1-n_\infty).
\end{eqnarray}
As it follows from Fig.~\ref{fig:abcd}(d), expression (\ref{eq:corre}) gives a  good agreement with the numerical results for overcritical domain.

Expressions (\ref{eq:jjfkdjwi}) or (\ref{eq:n-infty-n0 b}) allow us also to obtain the simple analytical expression of the form $B_c=B(\bar n,U,g)=0$ for the critical interface separating two different classes of NESS in the space of system parameters $(\bar n,U,g)$ (the transition phase diagram).  The two-domain structure becomes unstable when $n_\infty$ approaches to the equilibrium concentration $n_e\approx \bar{n}$, see Figs.~\ref{fig:abcd}(b). Setting $n_\infty\approx \bar n$ in Eq.~(\ref{eq:jjfkdjwi}) or Eq.~(\ref{eq:n-infty-n0 b}) we obtain approximate equation for critical surface $B_c$:
\begin{equation}\label{eq:vo2112f85J4K}
  g  \approx \frac{(1-U)(1-2\bar{n})}{4\bar{n}(1-\bar{n})-(1-U)},
\end{equation}
which gives more-or-less good agrement with the numerical results, see Fig.~\ref{fig:4b}.

Note that expressions $n_0=(1-U)/2$ and $n_{-1}+n_1=1$ are exact and satisfy Eqs.~(\ref{eq:n-infty-n0}). This follows from the equation for impurity site occupancy $ \partial_t n_0=J_{-1,0}-J_{0,1}=0$ that expresses $n_0$ in terms of gas states on impurity edges:
\begin{equation}\label{eq:DTno=0}
 n_0=\frac{1-U}{2}\left(1+\frac{n_{-1}+n_1 -1}{g(n_{-1}-n_1)+1}\right).
\end{equation}
This means that after the transition to the steady two-domain gas structure, the system possesses two invariants---impurity occupancy $n_0$ and the total occupation of its edges $n_{-1}+n_1=1$ which are insensitive to variations of the main system parameters, Fig.~\ref{fig:3a*}. Such a behavior of impurity edges can be considered as strong correlation of gas states at impurity edges, and the nonequilibrium transition can be characterized by switching between the diffusive skin effect and the edge-edge correlation one. Note that we exploit the rough model of asymmetric hopping transport. Our consideration is limited by the coordinate representation and loses  the spectral properties of the system which are basic in the description of many exotic topological phenomena in non-Hermitian systems, see e.g., \cite{Torres2018,torres2019perspective,song2019,yao2018,bergholtz2021,shen2018}.

The universal behavior of the local invariants  enables us to suppose that impurity state and correlated state of its edges  can behave as local first integrals, i.e., $n_0(t)=\text{const}$ and $ n_{+1}(t)+n_{-1}(t)=1$, after nonequilibrium phase transition ($g>g_c, U>U_c,\bar{n}>\bar{n}_c$) and, as a result, can become  insensitive to the noise of external field $\delta g(t)$ and to the fluctuations of the number of particles $\delta N(t)$ in the system. To qualitatively demonstrate the emergence of local first integrals we resort to the particular case of the time-dependent external non-conservative force $g(t)=g+\delta g(t)$.

\section{\label{sec:III-LIOMs}Emergent Local first integrals}
\begin{figure}
    \includegraphics[width=\columnwidth]{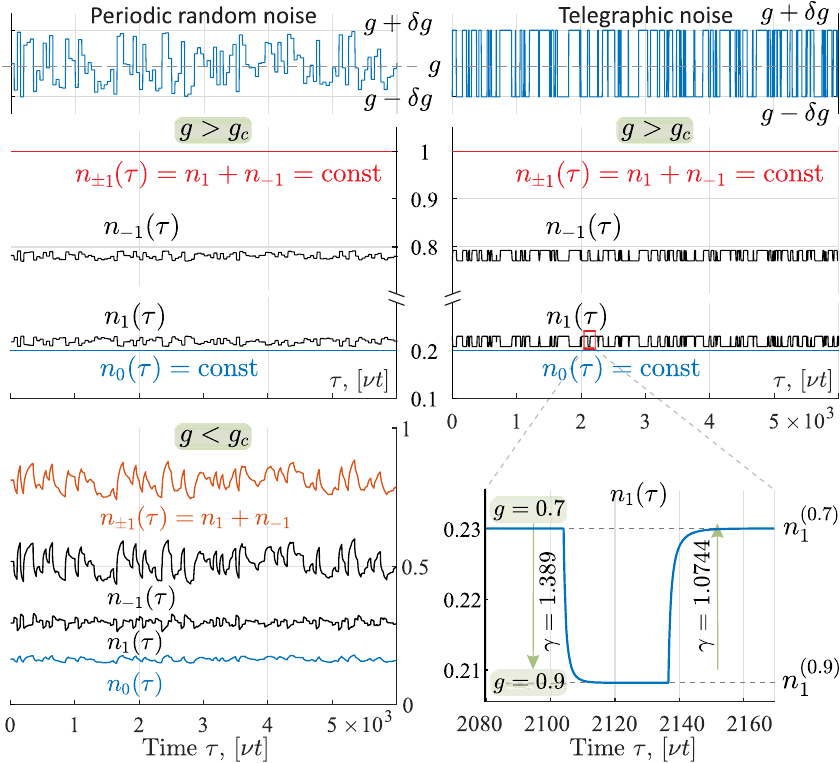}
    \caption{\label{fig:EIOM}Numerical illustration of the emergence of local invariants (local first integrals) $n_0(\tau)=(1-U)/2=\mathrm{const}$ and $n_{\pm1}(\tau)=n_1(\tau)+n_{-1}(\tau)\approx1$ after nonequilibrium transition at $g>g_c$ for the case of driving field noise $g(\tau)=g+\delta g(\tau)$.
    Exploited noise samples are shown at the top panel, both have switching frequency $\lambda=0.02$. For both kinds of noises the drive $g(t)$ fluctuates around $g=0.2$ (subcritical regime) or $g=0.8$ (overcritical regime) with maximum amplitude $|\delta g(t)|=0.1$. Here, ring length $2L=400\ell$, average density (filling fraction) $\bar n=0.3$, and $U=0.6$. Enlarged segment of $n_1(\tau)$ shows relaxation between $|g=0.7\rangle \rightleftarrows |g=0.9\rangle$ with dominant asymptotic behavior $\sim \e^{-\gamma\tau}$, implying decay rate $\gamma\gg\lambda$.
    The results shown are obtained based on the numerical solution of mean-field equations~(\ref{eq:nk})--(\ref{eq:Jk,k+1}) with $\delta g(\tau_k)$ being a sampled random realization of a stochastic process (either telegraphic or random periodic noise), see Appendix~\ref{sec:appendix_3}6 for details.}
\end{figure}
\begin{figure*}
  \includegraphics[width=.95\textwidth]{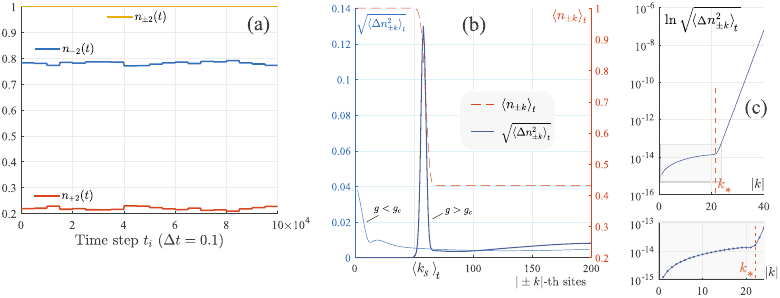}
  \caption{\label{fig:npmcum} Time invariance of quantity $n_{\pm k}=n_{-k} + n_{+k}\approx1$ against dynamic fluctuations of driving $g$ within $g>g_c$ domain. External drive $g(\tau)$ is governed by periodic random noise around the mean value $g=0.8>g_c$ as in Fig.~\ref{fig:EIOM} (see caption). In the subplot (a) the example of time behavior of $n_{\pm k}$ is shown for $k=2$. In the subplot (b) the averaging is performed over $6\times10^3\ [\nu t]$ during which $g(\tau)$ approximately reaches the mean value $g=0.8\pm0.01$. The quantity $n_{\pm k}(t)$ holds quasi-invariant within the interval from $k=1$ up to $\approx k_s$. As can be seen from (c) the invariance weakens slowly until certain $k_\ast$ which can be naturally associated with the size of obstacle complex. In the interval $k_\ast<k<{\langle k_s \rangle}_t$, the quasi-invariants $n_{\pm k}\approx1$ undergo  destruction and reach complete destruction at the domain wall position ${\langle k_s\rangle}_t$, as confirmed by numerical calculation of dispersion  ${\langle\Delta n^2_{\pm k}\rangle}_t={\left\langle (n_{\pm k}-{\langle n_{\pm k}\rangle}_t)^2 \right\rangle}_t\approx10^{-30}-10^{-9}$  for dense gas phase $0<k<40$, where ${\langle\dots\rangle}_t=\tfrac{1}{T}\sum_{t=0}^T\dots$ and symbol $\Delta$ is used (instead of $\delta$ in the main text) in order to denote numerically obtained values.
  One can say that obstacle complex, i.e., the domain $|k|<k_\ast$, is protected from $g$ noise-induced fluctuations.
  The results shown are obtained based on the numerical solution of mean-field equations~(\ref{eq:nk})--(\ref{eq:Jk,k+1}) with $\delta g(\tau_k)$ being a sampled random realization of the periodic random process, see Appendix~\ref{sec:appendix_3}6 for details.}
\end{figure*}
Generally, the problem of obtaining invariants and local first integrals  for the dynamical system (\ref{eq:nk})--(\ref{eq:Jk,k+1}) is of independent interest, see, e.g., \cite{arnold1992,GrobnerKnapp1967,Kryukov1995}. Here we qualitatively consider this question supplementing our  reasoning by numerical results. For the numerical simulations, we will consider  discrete changes of the field $g(t)$ in time, that is similar to telegraph-like processes.

To qualitatively illustrate emergence of local first integrals in the overcritical domain $(U>U_c,\bar{n}>\bar{n}_c,g>g_c)$, we resort to the simplest case when the field $g(t)$ jumps from $g=\text{const}>g_c$ to $g'=g+\delta g=\text{const}>g_c$ at time moment $t_0=0$ supposing that previously  the system is in steady state $\vec n^g=(n^g _{-L},\ldots,n^g _{L})^T$ (the superscript $T$ denotes transposition). In fact, we are interested in time-relaxation of the system from one steady state $\vec n^g$ (corresponding to field value $g$) to another one  $\vec n^{g'}$.
For this simple case, we show that occupation of the impurity site $n_0(t)=(1-U)/2$ and of its edges $n_1(t)+n_{-1}(t)=1$ can behave like local first integrals in the vicinity of noncritical point $\vec n^g$. Numerically, we demonstrate the emergence of these invariants as adiabatic ones at least for the system under the action of the time-fluctuating field $g(t)$.

For convenience, we rewrite the dynamical system (\ref{eq:nk})--(\ref{eq:Jk,k+1}) in the form
\begin{subequations}
\begin{equation}\label{eq:vecn}
\partial_t \vec n =  \vec {f}^{g'}(\vec n),
\end{equation}
\begin{equation}\label{eq:f}
 f^{g'}_i(\vec n)=J^{g'}_{i-1,i}-J^{g'}_{i,i+1},
\end{equation}
\end{subequations}
with initial condition
\begin{equation}\label{eq:ini}
\vec{n}\big|_{t=0}=\vec{n}^g.
\end{equation}
Here $\vec n=(n_{-L},\dots,n_{L})^T$,  $n_k$ is the mean occupation number of $k$th site, $\vec n^g$ is nonequilibrium steady state corresponding to field $g$, the bond flow  $J^{g'}_{i,i+1}$  given by expression (\ref{eq:4}) is written for field $g'=g+\delta g$, i.e., $\nu^{\pm}=\nu(1\pm g')$.

The formal solution of (\ref{eq:vecn}) can be represented in the Lie-series form, see, e.g., \cite{hydon2000,arnold1992},
\begin{equation}\label{eq:LieSolut}
  \vec n(t) = e^{tD^{g'}}\vec n\big|_{\vec n=\vec{n}^g},
\end{equation}
\begin{equation}\label{eq:LieD}
   D^{g'} = \vec {f}^{g'}(\vec n)\tfrac{\partial}{\partial \vec n}=\sum_i \left(J^{g'}_{i-1,i}-J^{g'}_{i,i+1}\right)\frac{\partial}{\partial n_i}.
\end{equation}
Taking into account that $\vec{f}^g(\vec{n}^g)=\vec{0}$ for steady state $\vec{n}^g$, and as a result $D^g \vec{n}\big|_{\vec{n}=\vec{n}^g}=0$, it is convenient to rewrite operator $D^{g'}$ in the form:
\begin{equation}\label{eq:DdD}
   D^{g'} =D^{g}+\delta D=\left( \vec{f}^g+\delta g\vec{\phi} \right)\frac{\partial}{\partial \vec n},
\end{equation}
where
\begin{equation}\label{eq:phi}
  \phi_i =  (1-2n_i-U\delta_{i,0})(n_{i-1} - n_{i+1}) + Un_i (\delta_{i,-1} - \delta_{i,1}).
\end{equation}
From (\ref{eq:phi}) and (\ref{eq:DdD}) it follows that
\begin{equation}\label{eq:Dg'no}
   D^{g'} n_0\big|_{\vec{n}=\vec{n}^g}=\phi_0\big|_{n^g_0=(1-U)/2}=0.
\end{equation}
This means  that $n^g_0=(1-U)/2$ behaves as local first integral at least in the vicinity of non critical point $\vec{n}=\vec{n}^g$ [$\vec{f}^{g'}(\vec{n}^g)\neq\vec{0}$], see Fig.~\ref{fig:EIOM}. Recall that $n_0^g=(1-U)/2$ is the exact steady state solution for impurity site only.\footnote{It would be desirable to show that $\exp (t D^{g'})n_0\big|_{\vec n^g} = n^g_0=(1-U)/2$. Note,  one can show that $\exp (t \delta D)n_0\big|_{\vec n^g} = \left[n_0+(1-U-2n_0) \xi (\vec n)\right]\big|_{\vec n^g}=n^g_0=(1-U)/2$ as well as $\exp (t  D^{g})\exp (t \delta D)n_0\big|_{\vec n^g} = n^g_0$, where $\xi$ is some function. However, to prove the general statement, using, for instance, Trotter's product formula, the exact steady state solution $\vec n^g$, i.e., for all the sites, is needed.}

Note that if $n_0=1-U-n_0=(1-U)/2$ were   integral of motion then quantity $n_{\pm1}(t) =n_{-1}(t)+n_1 (t)$ would be the integral of motion as well that immediately follows from equation for $n_0(t)$:
$$0=\partial_t n_0=J_{-1,0}-J_{0,1}=n_0(\nu^+ +\nu^-)[n_{-1}(t)+n_{1}(t)-1].$$

To qualitatively show that quantity $n_{\pm1}(t)\approx 1$ behaves like local invariant we use our approximate solution for steady state distribution $n_k^g$ with sharp step form (\ref{eq:step}). Action of Lie operator gives
\begin{gather}
D^{g'} (n_{-1}+n_1)\big|_{\vec{n}=\vec{n}^g} = \\
\left(\delta g/g \right)\left[ (n^g_{-1}+n^g_1)-(n^g_{-2}+n^g_2)\right]\Big|_{\substack{ n^g_{-1}=n^g_{-2}\approx1-n_\infty\\ n^g_1=n^g_2\approx n_\infty}} = 0\nonumber
\end{gather}
in the vicinity of  our approximate solution, see Fig.~\ref{fig:EIOM}. This approximation gives a series of  such quasi-invariants $n_{\pm k}(t)=n_{-k}(t)+n_k (t) =1$ which undergo destruction  near site $k=k_s$ corresponding to the position of interface between dense-depleted gas phases (domain wall). This is in qualitative agreement with numerical solution, Fig.~\ref{fig:npmcum}.
However,  numerical results suggest that  quantities  $n_{\pm k}$ behave like adiabatic invariants $\langle n_{\pm k}\rangle_t\approx1$ undergoing weakening as $k$ tends to $k=k_*<\langle k_s\rangle_t$ and complete destruction near characteristic wall position $\langle k_s\rangle_t$  as it  follows from the case of fluctuating field $g(t)$, see Fig.~\ref{fig:npmcum}. Here $\langle\ldots\rangle_t$ denotes time averaging.
The value $k_\ast$ can be related to the size of obstacle complex (the region $|k|<k_\ast$) which is protected from $g$-noise-induced fluctuations.
In contrast, the system loses the invariants and correlated time-behavior for sites near the impurity in the subcritical domain ($g(t)<g_c$, $U>U_c$, $\bar{n}>\bar{n}_c$). The state of the impurity and its  surroundings are sensitive to fluctuations of the driving field $g(t)$ as is shown in Figs.~\ref{fig:EIOM} and ~\ref{fig:npmcum}.

Note that the impurity filling conservation in time, $n_0=(1-U)/2=\text{const}$, requires equality of the incoming and outgoing currents at impurity, that means strong gas correlation at opposite its edges.
Synchronized behavior of the gas states at the opposite impurity edges $n_{-1}(t)+n_1(t)=1$ guarantees that incoming  and outgoing currents at the impurity are equal $J_{-1,0}(t)=J_{0,1}(t)=n_0\left[\nu^+(t)n_{-1}(t)-\nu^-(t)n_{1}(t)\right]
=\nu n_0\left[n_{-1}(t)- n_1(t)+g(t)\right]$, that is a necessary condition for impurity filling conservation. Such synchronized dynamics can be formally viewed as a manifestation of the edge correlation effect.

Thus, the nonlinear dynamic screening effect caused by gas flow scattering on the impurity  leads to the formation of gas dense phase near the impurity. This effect leads to decomposition of the system with emergence of invariant behavior of impurity state and strong correlated behavior of a gas on the impurity edges in overcritical domain.
The growth of dense gas phase near impurity can be speculatively considered as the growth of a nucleus around impurity serving as a nucleation centre the ``size'' of which is larger than the critical one that formally corresponds to the condition $U>U_c$. In addition, the boundary of nucleus (dense gas phase) being an additional scatterer  protects it's own center (impurity state) against fluctuations in a gas.

We can say that the interface appearing between  dense and depleted gas phases (the domain wall) results in protection of the  state of the ``impurity-dense gas'' complex   against fluctuations in the rest of the gas and against noise of external field.
We can associate such a decomposition with nonequilibrium protection effect by analogy with topological protection effect, and edge correlation one (see, e.g., \cite{garbe2023,minoguchi2023,cao2021,hu2022}).

\section{\label{IIIB}Spatial localization of induced fluctuations}
\begin{figure*}[t]
  \includegraphics[width=\textwidth]{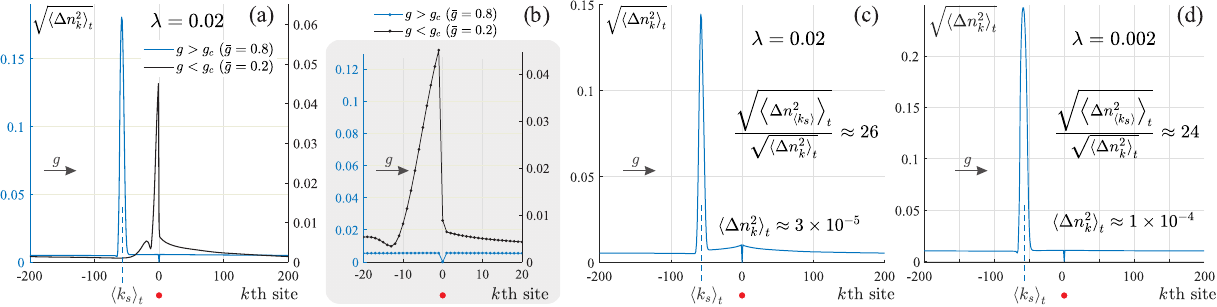}
  \caption{
  The square root dispersion ${\langle\Delta n^2_k\rangle}_t ^{1/2}={\left\langle (n_k-{\langle n_k\rangle}_t)^2 \right\rangle}_t ^{1/2}$ of the noise-induced fluctuations of the $k$th site occupation number (a) for periodic random noise with $\lambda=0.02$, such as in Figs.~\ref{fig:EIOM} and~\ref{fig:npmcum}(a), below $g_c$ (${\langle g \rangle}_t=0.2$) and above $g_c$ (${\langle g \rangle}_t=0.8$).
  (Symbol $\Delta$ is used, instead of $\delta$ in the main text, in order to denote numerically obtained values.)
  In the subcritical domain ($g<g_c$), the fluctuations induced by the multiplicative noise are mostly distributed near the impurity with the accumulation ahead of it, the site $k=0$. On the contrary, in overcritical domain ($g>g_c$), the noise-induced fluctuations are totally suppressed in the impurity, $k=0$, and strongly localized near the defect (the domain wall) position $\langle k_s \rangle_t$. This local enhancement of density-fluctuations intensity nearby the domain wall (phase coexistence layer) is caused by its back/forward trembling or, in other words, by noise-induced floating of the domain wall position with time, Fig.~\ref{fig:scheme}. (b) shows the zoomed in region from (a) near the impurity, $k\in[-20\ 20]$. (c) and (d) shows the same as (a) above $g_c$ (${\langle g \rangle}_t=0.8$, $\delta g=0.1$) but for telegraphic noises with $\lambda=0.02$ and $\lambda=0.002$, respectively. Other parameters: $2L=400\ell$, $\bar n=0.3$, and $U=0.6$.
  The results shown are obtained based on the numerical solution of mean-field equations~(\ref{eq:nk})--(\ref{eq:Jk,k+1}) with $\delta g(\tau_k)$ being a sampled random realization of a stochastic process (either telegraphic or random periodic noise), see Appendix~\ref{sec:appendix_3}6 for details.
  \label{fig:7}}
\end{figure*}

As we have seen, the reaction of the system to external noise $\delta g(t)$ below the phase transition ($g<g_c$) and above ($g>g_c$) is qualitatively different, see  Fig.~\ref{fig:EIOM}. Within the subcritical region, noise-induced fluctuations of density are distributed over all the system, reaching the maximum at impurity site, Fig.~\ref{fig:7},
see also Fig.~\ref{fig:npmcum}. Contrary, above the transition ($g>g_c$), the fluctuations are suppressed near impurity. In addition, the main part of the fluctuations in a gas are localized near nucleus boundary (domain wall), Figs.~\ref{fig:7} and~\ref{fig:npmcum}.

Here, we make an attempt to explain qualitatively the localization of induced fluctuations at the domain-wall position and to estimate the characteristic amplitude of density fluctuations within depleted gas phase and in the vicinity of nucleus boundary.

For the low-frequency external-field noise $\delta g(t)$, the system response  is determined by the system of nonlinear stochastic equations Eq.~(\ref{eq:8intext}) for the dynamics of mean occupation numbers $n_k(t)$ at lattice sites $k$, where $\nu^\pm(t)=\nu\pm \nu[g+\delta g(t)]$, see Appendix~\ref{sec:appendix_3}. We consider the perturbation of occupation numbers (concentration) at sites $n_k(t)=n^g_k+\delta n_k(t)$, caused by external noise $\delta g(t)$, close to NESS $n^g_k$ corresponding to field $g$ (supposing $|\delta g|\ll g$, $g+\delta g(t)>g_c$, and $|\delta n_k|\ll n^g_k$).
In this case, the kinetics of gas fluctuations $\delta n_k$ is governed by the complete system of Langevin equations (\ref{eq:DnDgA})--(\ref{eq:CorrDnA}) for each site. However, as we have seen, the steady-state concentration distribution $n^g_k$ in depleted phase behind the impurity ($k\geq1$) up to the position of domain wall ($k<-S$) can be treated as constant $n^g_k\approx n_\infty$, Eq.~(\ref{eq:step}).

We consider the perturbation of occupation numbers (concentration) at sites $n_k(t)=n^g_k+\delta n_k(t)$, caused by external noise $\delta g(t)$, close to NESS $n^g_k$ corresponding to field $g$ (supposing $|\delta g|\ll g$ and $|\delta n_k|\ll n^g_k$). As we have seen, the steady-state concentration distribution $n^g_k$ in depleted phase behind the impurity ($k\geq1$) up to the position of domain wall ($k<-S$) can be treated as constant $n^g_k\approx n_\infty$, Eq.~(\ref{eq:step}). 

Then, to estimate characteristic amplitude of fluctuations $\sqrt{\langle\delta n_k^2\rangle}$ in the depleted phase it is sufficient to consider behavior of concentration $n_1(t)=n^g_1+\delta n_1(t)$ at the nearest to the impurity site $k=1$. Stochastic equation for $n_1(t)$ reads:
\begin{eqnarray}\label{eq:equation}
    \dot n_1(t) &=& J_{0,1} - J_{1,2} = \nu^+n_0\nh_1(t) - \nu^- n_1(t)\nh_0\nonumber\\
    &-& [\nu^+n_1(t) \nh_2(t) - \nu^- n_2(t) \nh_1(t)],
\end{eqnarray}
where $\nu^\pm=\nu\pm\nu (g+\delta g(t))$, $\nh_{1(2)}=1-n_{1(2)}$, and $n_0$ is taken into account to be the  local first integral, $n_0=\nb_0=1-U-n_0=(1-U)/2$. Assuming that the behaviors of mean occupation numbers $n_1(t)=n_1^g+\delta n_1(t)$ and $n_2(t)=n_2^g+\delta n_2(t)$ at neighboring sites $1$ and $2$ do not differ significantly, $n_2(t)\approx n_1(t) = n^g_1+\delta n_1$, within linear approximation, we obtain closed Langevin equation on the deviation of mean occupation number $\delta n_1(t)$ from its steady-state value $n^g_1=n_\infty$ at the nearest to the impurity site:
\begin{equation}\label{eq:kljsdflksjd}
  \delta \dot n_1 \approx -\gamma \delta n_1 - A\delta g,
\end{equation}
where
\begin{equation}\label{eq:gamma}
  \gamma = 2\nu[n_0+g(1-2n_1)]= 2\nu[n_0+g(1-2n_\infty)],
\end{equation}
and
\begin{equation}\label{eq:A}
  A = -\nu[n_0 - 2n_1\nb_1]= \nu n_0(1-2n_\infty)/g.
\end{equation}
In the latter expression we have accounted for Eq.~(\ref{eq:n-infty-n0 b}).

The rate $\gamma$, Eq.~(\ref{eq:gamma}), is the rate of relaxation towards a  NESS $|g,U,\nb\rangle$ corresponding to the kink configuration determined by the external field $g>g_c$, impurity state $U>U_c$ ($n_0=(1-U)/2$), and mean gas concentration (or particle number) $\nb>\nb_c$.

It is worth to note that relaxation rate, Eq.~(\ref{eq:gamma}), depends on $g$ and $U$ only,
$\gamma = \gamma_{g,U}$,
and does not depend on mean concentration $\nb$. This means, that transition rates $\gamma _{1,2}$ and $\gamma _{2,1}$ between two NESSs, $|1\rangle$ and $|2\rangle$, with different $g_1\neq g_2$ and/or $U_1\neq U_2$, are different, $\gamma _{1,2}\neq \gamma _{2,1}$, see Sec.~\ref{sec:gamma} for details.

It is easy to write the solution of Eq.~(\ref{eq:kljsdflksjd}) and the expression for autocorrelation function or dispersion of fluctuations of the mean occupation number $\delta n_1$ close to the NESS $n^g_1 =n_\infty (g)$:
\begin{equation}\label{eq:sol}
  \delta n_1 = - \mathrm{e}^{-\gamma t} \displaystyle\int\limits_0^t \mathrm{d}t_1 \mathrm{e}^{\gamma t_1} A\delta g(t_1),
\end{equation}
\begin{equation}\label{eq:meandn}
  \left\langle \delta n_1^2 \right\rangle = \mathrm{e}^{-2\gamma t}\displaystyle\int \limits_0^t \mathrm{d}t_1\int\limits_0^t \mathrm{d}t_2\mathrm{e}^{\gamma(t_1+t_2)} A^2\langle \delta g(t_1)\delta g(t_2) \rangle.
\end{equation}
Here, we assume that at the initial moment the system was in a NESS $n_1(t=0) =n^g_1=n_\infty (g)$.

For simplicity, we consider the ordinary telegraphic process \cite{gardiner,bala}, see Fig.~\ref{fig:EIOM}, where $\delta g(t)$ takes two values, $\pm \delta g$, with amplitude $\delta g$ and switching rate $\lambda$, for which the correlation function takes the form $\langle \delta g(t_1) \delta g(t_2)\rangle = (\delta g)^2\displaystyle \exp(-2\lambda|t_1-t_2|)$, and expression (\ref{eq:meandn}) is written as $\left \langle \delta n_{1}^2 \right\rangle\approx (\delta g A)^2/[\gamma(\gamma-2\lambda)]$ at $t\rightarrow\infty$.

For the parameter values used in numerical solution (Figs.~\ref{fig:EIOM} and~\ref{fig:7}), $n_0 = 0.2$, $n_\infty \approx 0.22$, $g=0.8$, noise switching rate $\lambda \sim 2\times10^{-2}\nu$, and noise amplitude $\delta g \sim 10^{-1}$, we obtain the decay rate $\gamma \sim 1.3 \nu$ with Eq.~(\ref{eq:gamma}), that is consistent with numerical results, cf. inset in Fig.~\ref{fig:npmcum}. It is easy to notice, $\gamma \gg \lambda$.
In this case, dispersion Eq.~(\ref{eq:meandn}) takes the form
\begin{eqnarray}\label{eq:disp1}
  \left \langle \delta n_{1(\infty)}^2 \right\rangle \approx  \left( \frac{\delta g A}{ \gamma} \right)^2\nonumber\\
  = \left[ \frac{ n_0(1-2n_\infty)\delta g}{2g[n_0+g(1-2n_\infty)]} \right]^2 \propto 1.16\times10^{-4},
\end{eqnarray}
which also agrees well with numerical calculations for characteristic dispersion of concentration outside the impurity and domain wall, see Figs.~\ref{fig:7}. 

Obtained approximate results make sense only if $g\gg |\delta g|$  and $|g+\delta g(t)|>g_c $, and do not applicable for $g\approx 0$.
To consider fluctuations $\delta n$ near equilibrium steady state $n^\text{e}$, corresponding to $g=0$, so simple approximation is not enough.
The condition  $g\gg |\delta g|$ means that field noise $\delta g(t)$ does not change the direction of driving field.

\begin{figure}
  \includegraphics[width=.75\columnwidth]{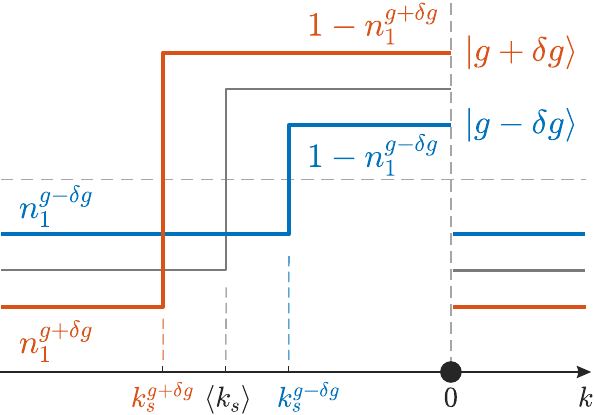}
    \caption{\label{fig:scheme}Schematic representation of $g$-noise-induced wall shifting approximation.}
\end{figure}

Let us estimate the dispersion of fluctuations around characteristic position of domain wall. Note, condition $\gamma \gg \lambda$ implies that the system asymptotically reaches its NESS over the time between two consecutive switches of the field $\delta g(t)$.
In other words, the system follows the field and adapts to its new values with a slight delay.
That facilitates the problem---assuming fluctuations in the system $\delta n_{1}(t)$ to be governed by the telegraphic process as well. The system stays on average (equally-likely) in two NESSs, $|g+\delta g\rangle$ and $|g-\delta g\rangle$ described by the kinks at the fields $g+\delta g$ and $g-\delta g$, see~Fig.~\ref{fig:scheme}.

The distribution function for $n_1$ can be written as follows
\begin{equation}\label{eq:PDF}
  \mathcal{P}(n_1) = \frac{1}{2}\delta (n_1-n^{g+\delta g}_1) + \frac{1}{2}\delta (n_1-n^{g-\delta g}_1),
\end{equation}
where $n^{g\pm \delta g}_1$ denotes steady-state values of $n_1$ at the field $g\pm \delta g$. Far away from the domain wall and impurity site, the dispersion of induced fluctuations of concentration takes the form
\begin{equation}\label{eq:dispn2} 
   \langle\delta n_1^2\rangle=\left\langle n_1^2 - \langle n_1\rangle^2\right\rangle = \frac{(n^{g-\delta g}_1-n^{g+\delta g}_1)^2}{4}.
\end{equation}
Here, $n_1^{g\pm\delta g} = n_\infty(g\pm\delta g)$,
where $n_\infty(g\pm\delta g)$ is given by Eq.~(\ref{eq:jjfkdjwi}). The explicit form of expression for $\langle\delta n^2_{\langle k_s\rangle}\rangle$ is a bit cumbersome and is presented in Appendix~\ref{sec:appendix_2}. However, in the limiting case $g\gg\delta g$, we can approximately write $n_\infty(g\pm\delta g)\approx n_\infty(g)\pm n_\infty'(g)\delta g$, where $n_\infty'(g)=\tfrac{\mathrm{d}}{\mathrm{d}g}n_\infty(g)$ and $n_\infty(g)$ is given by Eq.~(\ref{eq:jjfkdjwi}). Then,
$$\frac{1}{2}\left(n^{g-\delta g} - n^{g+\delta g}\right)\approx-n_\infty'\delta g
\approx \frac{n_0}{2g}\frac{(1-2n_\infty)\delta g}{n_0+g(1-2n_\infty)}$$
that expectedly leads to the same expression for $\langle\delta n_1^2\rangle$ as in Eq.~(\ref{eq:disp1}), obtained in a different manner. The direct evaluation of Eq.~(\ref{eq:dispn2}) for the given parameters yields $\langle\delta n_1^2\rangle \approx 1.44\times10^{-4}$, cf. Eq.~(\ref{eq:disp1}), which gives a good agreement with numerical calculations, Figs.~\ref{fig:7}(a) and \ref{fig:7}(c).
The mean position of the kink-wall (site's number) $\langle k_s\rangle\approx k^g_s= -S$, see Eq.~(\ref{eq:S}), and the kink height $\approx (1-n^g_1)$ are determined by the stationary value $n^g_1$ at the field $g$.
Taking into account that mean position $k^{g+\delta g}_s< \langle k_s \rangle <k^{g-\delta g}_s$, Fig.~\ref{fig:scheme}, we can roughly estimate the dispersion of fluctuations of site occupation nearby the characteristic position of domain wall, namely, at the site $\langle k_s \rangle$. The distribution function for $ n_{\langle k_s\rangle}$ can be represented as, see~Fig.~\ref{fig:scheme},
\begin{equation}\label{eq:Pnw}
  \mathcal{P}\!\left(n_{\langle k_s\rangle}\right)\! =\! \frac{1}{2}\delta(n_{\langle k_s\rangle}-(1-n^{g+\delta g}_1)) + \frac{1}{2}\delta(n_{\langle k_s\rangle}-n^{g-\delta g}_1),
\end{equation}
and the dispersion takes the form
\begin{equation}\label{eq:dispw}
\langle\delta n^2_{\langle k_s\rangle}\rangle=\left\langle n_{\langle k_s\rangle}^2 - \langle n_{\langle k_s\rangle}\rangle^2\right\rangle = \left(\frac{1}{2}-\langle n_1\rangle\right)^2,
\end{equation}
where $\langle n_1\rangle=(n^{g-\delta g}_1 + n^{g+\delta g}_1)/2$.
In the limit $g\gg\delta g$, $\sqrt{\langle \delta n_{\langle k_s\rangle}^2\rangle}\approx\tfrac{1}{2}(1-2n_\infty) + O(\delta g^2)$ characterizes the uncertainty of the center point along the vertical of the wall due to our approximate kink-solution with sharp form, Eq.~(\ref{eq:step}).
From (\ref{eq:dispw}) and (\ref{eq:dispn2}), one can obtain the estimate for characteristic enhancement of induced fluctuations of gas concentration in the vicinity of domain wall as compared to fluctuation distant from it:
\begin{equation}\label{eq:lljllk}
  \sqrt{\frac{\langle\delta n^2_{\langle k_s\rangle}\rangle}{\langle\delta n^2_1\rangle}} = \frac{1-n^{g-\delta g}_1 - n^{g+\delta g}_1}{n^{g-\delta g}_1-n^{g+\delta g}_1},
\end{equation}
In the limiting case $g\gg\delta g$,
\begin{equation}\label{eq:ratio}
    \sqrt{\frac{\langle\delta n^2_{\langle k_s\rangle}\rangle}{\langle\delta n^2_1\rangle}} \approx \frac{g}{\delta g} \frac{n_0+\left(1-2 n_{\infty}\right) g}{n_0},
\end{equation}
where $n_\infty=n_\infty(g,U)$ is given by Eq.~(\ref{eq:jjfkdjwi}), see Appendix~\ref{sec:appendix_2}.
The direct evaluation of Eq.~(\ref{eq:lljllk}) for the given parameters yields
$\sqrt{\langle\delta n^2_{\langle k_s\rangle}\rangle/\langle\delta n^2_1\rangle} \sim 23.57$, which shows more-or-less good consistence with numerical results, see Fig.~\ref{fig:7}(c).

The obtained results (\ref{eq:disp1}), (\ref{eq:dispn2}), and (\ref{eq:lljllk}) do not depend on the random switching rate $\lambda$ of the external field since the relaxation rate of the system $\gamma\gg\lambda$, as it was shown for the given system parameters as least, i.e.,  the system is also governed by the similar telegraphic process.
In this case, our simple qualitative estimates (\ref{eq:disp1}), (\ref{eq:dispn2}), (\ref{eq:dispw}), and (\ref{eq:lljllk}) give a good agreement with the direct numerical solution of the nonlinear stochastic Langevin equation based on Eq.~(\ref{eq:nk}) [Eq.~(\ref{eq:A6:nk})], with multiplicative noise given by the telegraphic process of the driving field $g(t)=g+\delta g(t)$.

Thus, it has been shown that the external field $g$-noise ($g>g_c$) induces density fluctuations only outside the impurity complex with their strong localization near the domain wall, Figs.~\ref{fig:7}(a)--(c). The latter is caused by the instability of the wall position, similar to the nucleus surface fluctuations during the first-order phase transitions, see, e.g., \cite{tosatti2014,khare1996}. In contrast, in the subcritical region ($g<g_c$), Fig.~\ref{fig:7}(a), the fluctuations are distributed throughout the system with a maximum near the impurity the magnitude of which is an order less than the one in supercritical region.

In the both cases, above and below critical point $g_c$, the most of the induced fluctuations are accumulated near lattice defects which act as the main scatterers of gas flow. 
In the first case ($g<g_c$), the induced fluctuations are concentrated only ahead of the impurity site (the obstacle), Figs.~\ref{fig:7}(a),(b).
The states of the sites behind the impurity are weakly affected by fluctuations, i.e., the scatterer suppresses fluctuations behind itself.
In the second case ($g>g_c$), corresponding to two-domain gas structure, the fluctuations are localized near the domain wall that can be described as the topological defect.
It is significant that gas fluctuations are totally suppressed inside the obstacle located behind the defect.
In other words, the tightly bound structure of the defect and the obstacle state, emerging after the phase transition, can be considered as an nonequilibrium quasiparticle similarly to one in liquid crystal \cite{Lev2002,Foffano2014-et}.
The role of the defect (domain wall) is to protect the  obstacle state from the external fluctuations, in our  particular case, from ones induced by a noise of the external driving field $g$.

\section{\label{sec:gamma}
Transition rates between different NESSs}

As it was mentioned above, the damping rate $\gamma$ of the fluctuations, induced by the external noise $\delta g(t)$ near NESS $|g\rangle$, depends on the magnitude of the carrier field $g$ and impurity state $n_0$ (or $U$), and it does not depend on mean particle concentration $\bar{n}$.
This means that transition rates $\gamma_{1,2}$ and $\gamma_{2,1}$ for transitions between two NESSs $|1\rangle\rightarrow |2\rangle$ and $|2\rangle\rightarrow |1\rangle$ with different $g_1\neq g_2$ and/or $U_1\neq U_2$  can be different, $\gamma _{1,2}\neq \gamma _{2,1}$.

Here we make simple analytical estimation of these rates comparing them with asymptotic form of relaxation time values obtained from numerical solution, and show that  $\gamma_{1,2}= \gamma _{2,1}$ if states $|1\rangle$ and $|2\rangle$ are different in field direction $\pm g$ that corresponds to transitions between steady states $|g\rangle\leftrightarrows|-g\rangle$.

\begin{figure}
    \includegraphics[width=.9\columnwidth]{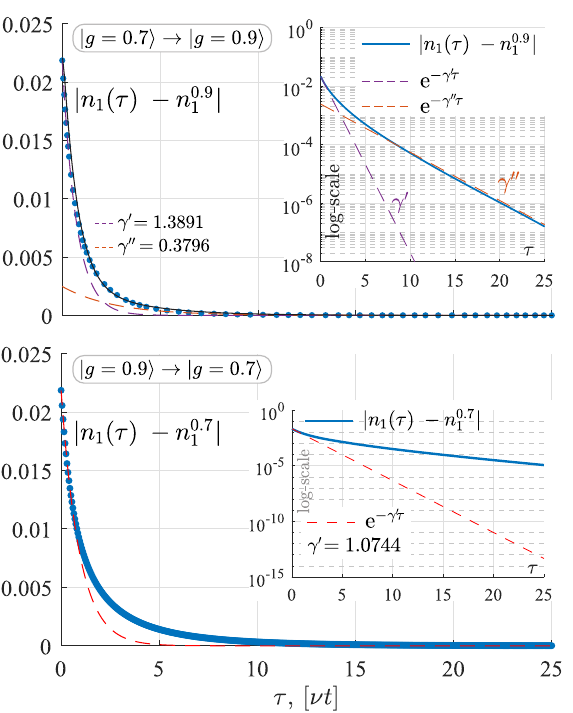}
    \caption{\label{fig:gamma07_09}
  Overcritical regime (within the domain $g>g_c$) Numerically obtained $\gamma$ for transitions $|g=0.7\rangle \rightleftarrows |g=0.9\rangle$. Generally, $n_1(\nu t)$ decays according to Eq.~(\ref{eq:n1asympth}), where exponential behavior is decomposed into a fast decay with $\gamma'$ at initial times and a slower decay with $\gamma''$ at moderate times, see figure insets; for estimate on upper plot (solid line) $b=0.0025$, $\alpha = 0.3$. As a rough estimate, we can use dominant (fastest) asymptotic behavior of transition rate $\gamma'$, and associate characteristic relaxation time with $\tau=\tfrac{1}{\gamma'}$.
  The results shown are obtained based on the numerical solution of mean-field equations~(\ref{eq:nk})--(\ref{eq:Jk,k+1}), see Appendix~\ref{sec:appendix_3}6 for details.}
\end{figure}
We consider relaxation of the system to its NESS $|g'\rangle$ supposing that initially, i.e., at $t=0$, the system was in the NESS $|g\rangle$, i.e., the driving field $g$ abruptly (non-adiabatically) changes its value form $g$ to $g'$ at $t=0$.

To qualitatively demonstrate that $\gamma_{g',g} \neq \gamma_{g,g'}$ we again consider the time behavior of mean occupation number $n_1(t)$  only for the nearest to the impurity site $k=1$:
\begin{eqnarray}\label{1}
  \dot{n}_1 &=& \nu^+_{g'}n_0\nh_1 - \nu^-_{g'}n_1\nh_0
  - [\nu^+_{g'}n_1\nh_2 - \nu^-_{g'}n_2\nh_1],
\end{eqnarray}
with initial condition $n_1|_{t=0}=n_1^g$. Here $\nu^{\pm}_{g'}=\nu (1\pm g')$ and $n_1^g$ is steady state of $n_1$ corresponding to the field magnitude $g$.

\begin{figure}
    \includegraphics[width=\columnwidth]{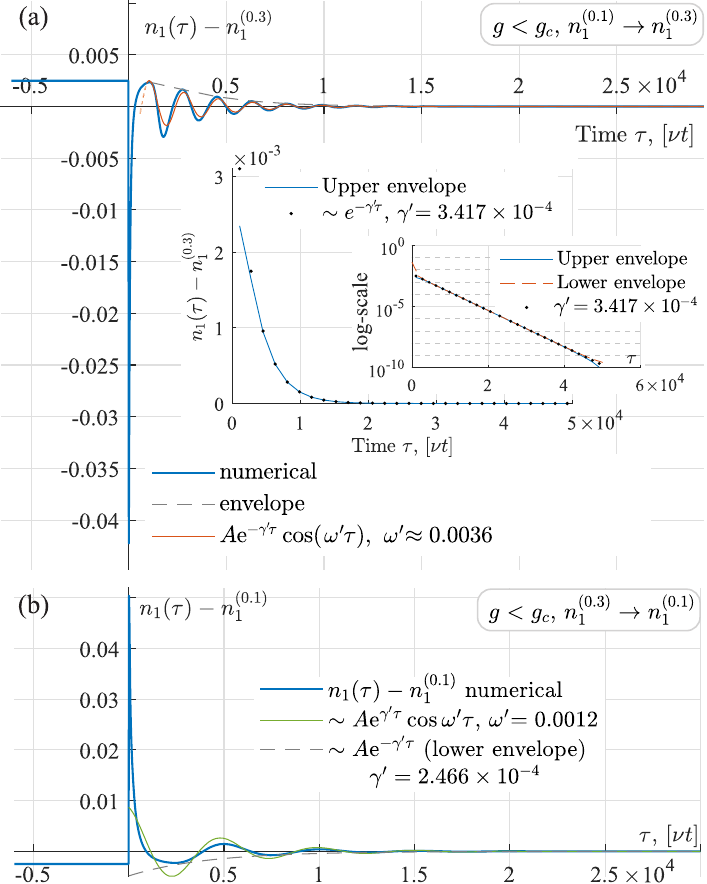}
    \caption{\label{fig:gllg_c}
    Subcritical regime (within the domain $g<g_c$):
    (a) Numerically obtained decay rate $\gamma'$ for transition from $g=0.1$ to $g=0.3$. $n_1(\tau)$ decays as $\sim A\e^{-\gamma'\tau}\cos(\omega'\tau)$ with exponent describing the envelope behavior, and $A=n_1^{(0.3)}-n_1^{(0.1)}=-0.0025$.
    (b) The same for reverse transition. $n_1(\tau)$ decays as $\sim A\e^{-\gamma'\tau}\cos(\omega'\tau)$, with $A=n_1^{(0.1)}-n_1^{(0.3)}=0.0025$ and exponent describing corresponding envelope behavior.
    The results shown are obtained based on the numerical solution of mean-field equations~(\ref{eq:nk})--(\ref{eq:Jk,k+1}), see Appendix~\ref{sec:appendix_3}6 for details.}
\end{figure}

The numerical results, see Figs.~\ref{fig:gamma07_09} and \ref{fig:gllg_c}, demonstrate
two-stage relaxation mechanism for the transition $n_1^g
\rightarrow n_1^{g'}$: the fast relaxation $\gamma'_{g,g'}$ prevails for initial times, and slow one $\gamma''_{g,g'}$ does for long time scales, $\gamma'_{g,g'}\gg \gamma''_{g,g'}$:
\begin{equation}\label{eq:n1asympth}
  |n_1(t) - n_1^{g'}|\approx |n_1^g-n_1^{g'}|\exp(-\gamma'_{g,g'}t)+b t^{\alpha}\exp(-\gamma''_{g,g'}t)
\end{equation}
obtained from numerical solution of complete system of
nonlinear equations for $n_k (t)$, cf.~Fig.~\ref{fig:gamma07_09}.
The system transition to new steady state is collective processes, and is described by collective spectrum of relaxation rates (at least in the linear approximation ), as a result, the relaxation for each site is determined by the set of different relaxation times.
Above, we have estimated the relaxation rate $\gamma$ to the steady state $|g'\rangle$, Eq.~(\ref{eq:gamma}), that is obtained in the linear approximation for a small initial perturbation of this state, and as a result, is irrelevant to the initial state.

To take into account the initial state $|g\rangle$, and to qualitatively estimate the fast relaxation rate we consider the decay of the state $|g\rangle$ caused by an abrupt change of the driving field value from $g$ to $g'$.

We estimate the characteristic relaxation time $\tau = \tau_{g,g'}$ corresponding to the transition $n_1^g\rightarrow n_1^{g'}$ by the approximate formula
\begin{equation}\label{eq:1}
  \dot{n}_1|_{t=0} \approx \frac{1}{\tau}\displaystyle\int\limits_0^\tau \dot{n}_1\ \mathrm{d}t \approx \frac{n_1^g-n_1^{g'}}{\tau},
\end{equation}
formally supposing that mean rate of the transition $|g\rangle \rightarrow |g'\rangle$ coincides with the initial rate of the decay of state $|g\rangle$. We also assume and that $n_1(t)$ has decreasing behaviour and $n_1(\tau)\approx n_1^{g'}$. Expression for $\dot{n}_1|_{t=0}$ follows from (\ref{1}):
\begin{equation}\label{eq:(1.5)}
  \dot{n}_1|_{t=0} = \nu^+_{g'}n_0^g\nh_1^g - \nu^-_{g'}n_1^g\nh_0^g- \left(\nu^+_{g'}n_1^g\nh_2^g - \nu^-_{g'}n_2^g\nh_1^g\right).
\end{equation}
This expression can be simplified by representing acting field as $g'=g+\Delta g$ (where $\Delta g= g'-g$) and taking into account that  $n_1^g$, $n_2^g$, $n_0^g=n_0=\overline{n}_0=(1-U)/2$ are steady states corresponding to field $g$:
\begin{equation}\label{eq:(2.5)}
  \dot n_1|_{t=0} =\nu\Delta g\left[n_0 - (n_1^g\nh_2^g + n_2^g\nh_1^g)\right].
\end{equation}
From (\ref{1}) and (\ref{eq:(2.5)}), we obtain characteristic  time of transition from the state $|g\rangle$ to the state $|g'\rangle$:
\begin{equation}\label{eq:(2.6)}
  \tau_{g,g'} = \nu^{-1}\frac{n_1^{g'}-n_1^g}{g'-g}\left[n_0 - (n_1^g\nh_2^g + n_2^g\nh_1^g)\right]^{-1}.
\end{equation}
Taking into account the fact that $n_1^g\approx n_2^g$ and Eq.~(\ref{eq:n-infty-n0 b}) we can write the rough estimation for transition rate $\gamma_{g,g'}\sim\tau_{g,g'}^{-1}$ between two NESSs,
\begin{equation}\label{eq:(3.7)}
  \gamma_{g,g'} \approx -\nu\left(\frac{n_1^{g'}-n_1^g}{g'-g}\right)^{-1}\left[\frac{n_0}{g}\left(1-2n_1^g\right)\right].
\end{equation}
It directly follows from Eq.~(\ref{eq:(3.7)}) that $\gamma_{g,g'}\neq\gamma_{g',g}$:
\begin{equation}\label{eq:(3.8)}
  \frac{\gamma_{g,g'}}{\gamma_{g',g}}\approx \frac{g'}{g} \frac{1-2n_1^g}{1-2n_1^{g'}}.
\end{equation}
The ratio $\tfrac{\gamma_{g,g'}}{\gamma_{g',g}}>1$ for transitions with $g'>g$ as it follows from Fig.~\ref{G(x,y)G(y,x)}.
The estimated values of $\gamma_{g,g'}$ are in a  agreement with one of $\gamma'_{g,g'}$ obtained numerically that describes the fast relaxation corresponding to main asymptotics of Eq.~(\ref{eq:n1asympth}), Fig.~\ref{G(x,y)G(y,x)}.
\begin{figure}[t]
    \includegraphics[width=.9\columnwidth]{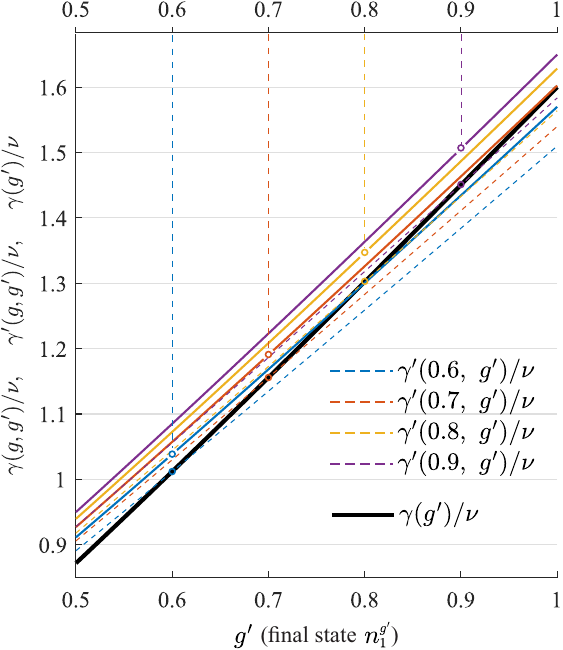}
    \caption{\label{G(x,y)G(y,x)} Transition rate $\gamma'(g,g')$ (numerical) corresponding to the fast-time part of relaxation from $|g\rangle \rightarrow |g'\rangle$, see Fig.~\ref{fig:gamma07_09}, solid lines plotted for different values of initial field  $g=0.6, 0.7, 0.8, 0.9$.  The transition rate $\gamma(g,g')$, Eq.~(\ref{eq:(3.7)}), is estimated as the decay rate of the state $|g\rangle$ at sudden field switching  $g\rightarrow g'$, dashed lines for the same parameters.
    The order of plotted lines, both solid and dashed, corresponds to the descending value of $g$, in accordance with the legend: lines higher on the plot correspond to larger values of $g$.
    Empty circles correspond to $g\equiv g'$ case. $\gamma (g')$ is relaxation rate to steady state $|g'\rangle$ regardless of an initial state $|g\rangle$, estimated by Eq.~(\ref{eq:gamma}), thick solid line.
    The results shown are obtained based on the numerical solution of mean-field equations~(\ref{eq:nk})--(\ref{eq:Jk,k+1}), see Appendix~\ref{sec:appendix_3}6 for details.}
\end{figure}
As an example, for transitions $|g=0.7\rangle \rightleftarrows |g'=0.9\rangle$ at given $U=0.6$, $\bar n=0.3$, the estimated values are $\gamma_{g,g'}=1.286\nu$ (numerically $\gamma'_{g,g'}=1.389\nu$) and $\gamma_{g',g}=1.089\nu$ (numerically $\gamma'_{g',g}=1.074\nu$).\!\footnote{Note that the approximation  Eq.~(\ref{eq:1}) can be applied to any site $k$: $\tau_k \approx (n^g_k-n^{g'}_k)/\dot n_k|_{t=0}$. The fast relaxation stage prevails for the regions with strong inhomogeneous distribution $n^g_k$ of initial state.}

However, $\gamma_{g,g'}=\gamma_{g',g}$ can be realized for transitions between steady states with different directions of the field $g$, i.e., $|g\rangle\leftrightarrows|-g\rangle$.
This is intuitively clear as changing of  the field sign to opposite leads to the similar kink-form NESS with gas concentration distribution $n_k^{-g}=n_{-k}^g$ obtained from initial concentration $n_k^g$ by inversion with respect to impurity site $k=0$. Note, that the states $|\pm g\rangle$ correspond to the states with different sign of the winding number naturally appearing for ASEP on the ring, see, e.g.,~\cite{popkov2011,prolhac2013}.

The situation is different for the transitions between steady states below critical point ($g, g'<g_c$). The  dynamics of  $n_1(t)$ during the transition $|g\rangle \rightarrow |g'\rangle$ is shown on Figs.~\ref{fig:gllg_c}(a) and \ref{fig:gllg_c}(b).
The evolution of this transition in time has a damped oscillating form.
In particular, the main asymptotics for $n_1(t)$ behaves like   $\sim\exp(-\gamma'_{g,g'}+i\omega'_{g,g'})t + \mathrm{c.c.}$, see Figs.~\ref{fig:gllg_c}.
The behavior of $n_\infty (t)$, and $\overline{J}(t)$ has similar oscillating character as well.
This indicates that eigenvalues of the system of the equations (at least for the linearized one) have complex values  in contrast to the case $g>g_c$, where eigenvalues are real.
The nature of these oscillations is in the generation of shock wave (solitary wave) on the one of impurity edges, at non-adiabatic switching of the field from $g$ to $g'$, which many times runs  around the ring.
More precisely, the system rebuilds  from one steady state to another in portions by sequential generation of solitary wave at the back side of impurity and by its traveling around the ring.
In the case of $g>g_c$, one solitary wave is enough to rearrange from one kink state to another.
In the latter case, the transmission of a solitary dissipative wave through the dense phase is hindered due to prohibited diffusion of particles in this localized region.
This is another interesting manifestation of  protection phenomenon  concerning the dynamics of steady state formation. Sudden changes of the driving field, either in subcritical or supercritical regime, always give rise to the creation of a ``quasi-particle'', namely, the generation and traveling dissipative soliton in the wake region of impurity, see \cite{pre2018}.
The steady state is established as a result of solitary wave run around the system.
While below critical point, such a soliton can make several turns over the ring passing through the impurity again and again, above the transition point we observe only one turn, so that the transmission of the quasi-particle through impurity-nucleus complex appears to be prohibited.
This forbidden transport of dissipative solitons can be interpreted physically by the absence of diffusion flow within the flat density plateau of the stratum region: locally suppressed diffusion (dissipative) processes makes the existence of dissipation assisted quasi-particles impossible. Thus, the total diffusion flow over the ring vanishes as well. The impurity becomes protected from the dissipative solitary perturbations traveling over the ring. Similar effect of the forbidden diffusion of quasi-particles was found in \cite{kawabata2023} under entanglement phase transition induced by the non-Hermitian skin effect.

It is necessary to emphasize that magnitudes of damping rates $\gamma_{g,g'}$ of inter-NESS transitions before ($\{g_1,g_1'\}<g_c$) and after ($\{g_2,g_2'\}>g_c$) phase transition can differ  significantly so that $\gamma_{g_1,g_1'}\ll\gamma_{g_2,g_2'}$.
In the particular case $\gamma'_{(0.1),(0.3)}/\gamma'_{(0.7),(0.9)} \propto 10^{-4}$.
This means strongly correlated behavior of the system with a domain structure above the critical point $g_c$.

Note that we can consider the $g$-field environment as a second ``reservoir'' for gas system in addition to  the thermal bath that initiates usual particle diffusion. For example, when adsorbed atoms  interact simultaneously  with thermal phonons of a solid-state substrate, and with the fluctuations of the external electromagnetic field. The fact that the transition rates between two steady states are different,    $\gamma_{g,g'}\neq\gamma_{g',g}$, could mean that the additional ``heat'' $Q\propto \ln\gamma_{g,g'}/\gamma_{g',g}$ can be transferred to $g$-field ``reservoir'' during such switching, see, e.g., \cite{Jarzynski2016,hatano2001}. However, this question requires more careful and detailed consideration that goes beyond our approximations.

\section{\label{sec:conclusion}Conclusions and discussion}
In this work, we have focused on some features of the nonequilibrium transition leading to the formation of nonlinear steady-state structures. The corresponding NESS configuration results from the gas flow scattering on a partially penetrable obstacle. The occurrence of nonlinear structures is provoked by the blockade effect in a gas with short-range repulsion. It manifests itself in the formation of a two-domain gas structure with a sharp boundary of dense-phase (kink-form profile)  ahead of the obstacle. This transition takes place when the values of mean gas concentration, driving field, and obstacle potential exceed certain critical values $(\bar{n}_c, g_c, U_c)$.

The main result of this work is that such transitions can be accompanied by the emergence of local invariants (local first integrals) which indicate the spatial system decomposition. The first invariant corresponds to the conservation of the number of gas particles inside the obstacle. This invariant is insensitive to a further increase or changes in the mean gas concentration or in the driving field. As a result, within this overcritical region,  the obstacle state becomes  robust against  temporal fluctuations or noise of the external driving field. At the same time, the rest of the system remains dependent on the total particle number and the external field. The necessary condition for such a decomposition is that the total current flowing through the obstacle boundary should be equal to zero, despite the fact that the local current fluctuates in time all over the system as well as inside the obstacle. This means synchronization between currents or gas states at different points of the obstacle boundary. Such a synchronized dynamics is described by the second local invariant that  can be viewed as manifestation of the edges correlation effect.
As a  matter of fact,  the rest of the system serves as a reservoir  for the obstacle that ensures the particle number conservation in this region as main parameters driving the system change.

Another interesting effect is the spatial localization of the gas fluctuations induced by the external drive noise in the overcritical regime. The domain wall between the dense and depleted gas phases as if captures or accumulates the fluctuations in the system, thereby protects the obstacle state against them. The concentration of induced fluctuations near the domain wall is caused by its  instability  under the action of the external noise. In some sense, this is similar to the growth of a nucleus at the first order phase transition, where the fluctuations of the nucleus surface are due to the active adsorbing and desorbing processes. The formation of the dense gas phase near the obstacle can be viewed  as the nucleus growth adjusting to the obstacle as to the nucleation center whose size is larger than the critical one ($U>U_c$). One can consider this effect as the manifestation of the nonequilibrium protection effect when  the domain wall serves as an additional scatterer and protects the state of the dense gas phase and the obstacle state.
Note that the similar effect can take place not only for a single obstacle but also for a finite-sized cluster of obstacles at least in the 1D case \cite{winpr}.

To demonstrate a possibility of these effects we consider the quasi-1D driven lattice gas with impurity sites as a limiting case of 2D or 3D  gas doped by impurity particles in a narrow channel. We resort to the simplest case of ring geometry and one  partially penetrable obstacle as a site with the reduced  mean number of  available vacancies $(1-U)$ that corresponds to the channel cell partially occupied by impurity particles with mean concentration $U$. To to describe the nonequilibrium steady-states and long-time fluctuations near them  we use coarse-grained description based on the combination of local equilibrium approach and mean-field approximation.

\medskip
In this setting, we found the form of phase diagram, i.e., the critical surface $(\bar{n}_c, g_c, U_c)$, for the transition to the two-domain gas structure with dense gas phase attached to the impurity. The transition is accompanied by the onset of two local invariants in relation to the external drive $g$, and gas concentration $\bar{n}$. The first one is the invariant state of the impurity site $n_0=(1-U)/2$ corresponding to its half-filling, regardless of the densities and sizes of two gas domains, determined by $\bar{n}$, $g$, and $U$.
Alone with integral order parameter, e.g., one describing enhancement of mean interparticle correlations, Fig.~\ref{fig:abcd}(d), the saturation of the impurity site, $n_0=(1-U)/2$, can serve as the local order parameter for this nonequilibrium transition.
Note that the impurity site saturation up to the finite value less than unity (half-filling) also means that the full blockade regime in the system can never be reached, except for the case of fully unpenetrable impurity site $U=1$.

Another emerging local invariant determines the strongly correlated behavior of the gas states at the opposite edges of the impurity and correspond at least to the adiabatic invariance of the total occupation of impurity edges (adjacent sites), $n_{-1}(t)+n_1(t)=1$. This leads to the equality of the incoming and outgoing currents at the impurity ensuring the particle conservation at the impurity site $n_0(t)=\text{const}$. Such a strong synchronized dynamics of currents at its opposite edges can be viewed as manifestation of the edge correlation effect. Note that this nonequilibrium transition can be characterized as the transformation of  the diffusive skin effect \cite{shankar2022,Liu2023} into the effect similar to the edge correlation one.

Our model have enabled the analysis of long-time-scale fluctuations and their spatial distribution near nonequilibrium steady states.
This concerns both thermal fluctuations as well as gas fluctuations induced by the noise of the field. To demonstrate that the dense gas phase protects the impurity complex from the external fluctuations, we consider the particular case of induced gas fluctuations provoked by the low-frequency noise of external driving field. $g(t)$ , Fig.~\ref{fig:EIOM}. The external field noise $g(t)$ at $g>g_c$ was shown to induce density long-time fluctuations only outside the impurity complex with their strong localization near the domain wall. The latter is caused by the fluctuations of the wall position. In contrast, in the subcritical region $\left(g<g_c\right)$, the fluctuations are distributed uniformly throughout the system with maximum localization near the impurity, cf.~Figs.~\ref{fig:npmcum} and~\ref{fig:7}. It is also shown that the characteristic relaxation times of the system to the nonequilibrium stationary state above the threshold are at least an order of magnitude higher than those below the transition. Note that the relaxation rates of the linearized dynamical system are complex below the threshold and are real above it.

The rate of transitions between two different NESSs corresponding to two different values of $g$ are different, except for the case where they are distinguished by the direction of non-conservative driving field $g$, or by the sign of their winding number, which is similar to non-Hermitian systems, see, e.g., \cite{shankar2022,bergholtz2021,gong2018}. Relaxation from one NESS to another is governed by the one-step or multi-step mechanism caused by the sequential generation of shock waves at one impurity edge. In addition, the transition rates demonstrate two time-scales of relaxation to a final state. 

\medskip
There are variety of 1D or quasi-1D models describing the nonequilibrium transition with the onset of domain structure caused by an obstacle in a drifting gas. These models differ by the manner of the implementation of an obstacle. The classical blockage 1D-lattice gas models associate the obstacle with defect bonds with modified hopping rates, often reduced rates \cite{janowski1992,janowsky1994,JanowskyLebowitz2000,costin2012,kolomeisky_1998,GREULICH20081972,Greulich_2008}. Defects are   assumed to be as static. Another approach is to associate the obstacles with the particles of the second gas component, impurities, usually assumed to be mobile. This class of models is related to the quasi-1D lattice models corresponding to the limiting case of a narrow channel. These models adequately describe the behavior  of colloidal, impurity, tracer particles, as well traffic jam effect in confined (narrow channel) geometry \cite{korniss_1999,benichou_2018,Miron_2020,miron2021,lobaskin2024,TabatabaeiSchutz2006,slanina2023}. Our model suggests that the second gas component is static, and can be referred to as the quasi-1D driven lattice gas with impurity site. Naturally that some our results reproduce in part the results obtained in other blockage models. In particular, the independence  of the gas concentration $n_\infty$ far from obstacle on the total particle number (or mean gas concentration $\bar n$), as well as the existence of several system phase states in the low and high density cases, cf. \cite{janowski1992,janowsky1994,JanowskyLebowitz2000,costin2012, mallick2015,lobaskin2022,lobaskin2023,lobaskin2024, kolomeisky_1998,GREULICH20081972,Greulich_2008}.
The  behaviors of critical transition curves in the classical slow bond model and in our ``impurity site'' one are in agreement at relatively weak field $g<1$; the phase boundary curves on the 2D coordinate plane: gas concentration and corresponding obstacle parameter, i.e.,  $(\bar n,\ \text{\textit{slow-bond rate/impurity-site vacancies}})$,
e.g., cf. Fig.~\ref{fig:4b}(a) and Fig.~18 in Ref.~\cite{mallick2015}, see also \cite{janowski1992,janowsky1994,JanowskyLebowitz2000,costin2012,mallick2015}.
The case of strong field $g\gg1$ assumes the high magnitude of gas current flowing along the chain, that can result in the nonequilibrium transition  induced by local gas fluctuations   without any obstacle \cite{schutzPTCP2001,bodineau2005,espigares2013,hurtado2011}. This case is not considered in the present paper.
Our results partially echo those obtained in recent works \cite{lobaskin2024,lobaskin2022} for the case of driven two-component system of  a mobile heavy particle (free defect) and normal gas component with reduced transition rate of their mutual inter-site exchange.
The first it is the fact that the system is practically insensitive to  the obstacle (free defect), the effects of which is local for subcritical regime.
As was shown in \cite{lobaskin2024}, the  change in time of defect position (its local jump) generates density wave going around the system. In our case, the static obstacle generates similar density waves at  jump in time of the value of external driving field $g$. 
In \cite{lobaskin2024}, it seems for the first time,  the emergence of global coherent effect have been shown at system transition to   the domain structure (shock phase), when local current fluctuations of   the defect particle  induce  long-range current fluctuations of the order of system size.
Authors of \cite{lobaskin2024} mainly focus on the value of the deviation of time- and space-averaged (over all the bonds) current   from its mean value based on the long time limit for large deviation function,\!\footnote{The dynamical phase transitions associated with time-integrated observables, in particular, fluctuations of total current through the system,
occurring in diffusive (ASEP) systems were extensively studied within the so-called macroscopic fluctuation theory \cite{Bertini2015-mf,Bertini2002-oj,Bertini2009-an,Bertini2007-vc,Bertini2005-ig,Bertini2001-mh}.} and also describe the discontinuous change in scaling behavior of diffusive constant, as of order parameter, at this nonequilibrium phase transition. In contrast, we stay in the framework of Langevin equations for the long-time- and spatial-fluctuations of gas concentration  near its spatially inhomogeneous nonequilibrium steady state to consider the local density deviations, that enables us to qualitatively explain their spatial localization near the domain wall.
One of the important deficiencies of our model is neglecting the fast processes in a gas.
This, in turn, leads to neglecting of short-range correlations between gas particles, see, e.g., \cite{Lukyanets2010,pre2017,pre2018}, and a possibility of nonequilibrium phase transition induced by local gas fluctuations without any obstacles \cite{schutzPTCP2001,bodineau2005,espigares2013,hurtado2011}, that is significant at high enough field $g$.

\medskip
The nonequilibrium transition with the formation of nonlinear structures like stratum, and effects like protection and edge synchronization  ones can take place in the case of gas flow scattering on a finite cluster of impurities. Note that the gas flow scattering on an impurity cluster and the total drag force acting on the cluster significantly depend on its structure factor. As was shown for 2D case, scattering and the drag force can be enhanced by the decomposition of a solid obstacle into fragments or a sparse cluster of impurities. This enhancement is due to the collective scattering effect implying the onset of the common wake around impurities, and it is more efficient for disordered clusters as compared to the regular ones, see, e.g., \cite{pre2018}.
At the same time, the mechanism of gas scattering determines the features of induced nonequilibrium (flow- or wake-mediated) inter-impurity interaction \cite{dzubiella2003,wulfert2017,pre2017,EPJST2013} which usually belongs to the non-reciprocal type \cite{sasa2006,ivlev2015}.
Crossover between linear and nonlinear regimes of dynamic screening can drastically change this interaction. In particular, the formation of a common coat of gas density perturbation accompanied by the blockade effect around impurities can result in their dissipative binding \cite{oshanin2011,vasilyev_2017}.
In the case of finite clusters, the collective scattering effect can lead to the shift of critical points of the nonequilibrium transition, for example, to the reduction of the critical value of driving field $g_c$. The protection effect can manifest itself in suppressing flow-induced correlations between obstacles after the nonequilibrium transition to the domain gas structure, at least in 1D \cite{winpr}.
The quasi-one-dimensional domain gas structures can serve as an example of nonequilibrium atomic-wire structure, whose shape is controlled by the external drive. This might be realized by the wind effect \cite{preis2021} for adsorbed atoms on the anisotropic or stepped solid-state substrate.
It would be interesting to consider a possibility for analogous system decomposition, boundary current synchronization, and protection  effect in 2D and 3D, i.e., when the wake around a cluster of obstacles protects its internal state or internal dynamics against external fluctuations or perturbations.

\acknowledgments{We are grateful to B.I.~Lev, I.S.~Gandzha, A.V.~Semenov, and Ch.~Riemann for helpful discussions, comments on the manuscript, and heartfelt support. Authors express sincere gratitude to the Krzysztof Skubiszewski Foundation for financial support as part of the programme ``Urgent Aid for the Preservation of the Research Capabilities of Ukraine''.}

\appendix

\section{\label{sec:appendix_3}Model}

In this section we describe the main approximations used in our model. The model is based on the combination of the local equilibrium approach, and the mean field approximation.  Firstly, we shortly discuss the local equilibrium approach that is based on the existence of different time-scales; the Langevin equation describing the  nonequilibrium fluctuations  for long time-scales, neglecting fast processes. Farther, we consider the mean field approximation in the framework of local equilibrium approach to describe nonequilibrium steady states and fluctuation near them. In the end, we briefly discuss the case of the narrow channel for  driven lattice gas doped by impurity (heavy) particles, and its reduction to  quasi-1D lattice.

\textbf{1. The Langevin equation for nonequilibrium hopping system.}
We consider a periodic potential for particles, for example, a potential relief of solid state substrate for adsorbed atoms (adatoms).
The potential relief forms the periodical lattice whose sites correspond to the potential relief minima. 
Adatoms spend most of time $\tau_l$ at the minima of relief and occasionally jump (overbarrier transition) to other minima due to thermal vibrations of the substrate atoms, that is possible for relatively small temperature.
We also suppose that only one adatom can be trapped by the minimum of potential relief due to strong short-range repulsion between adatoms, i.e., the  state of site $i$ can be described by its occupation number $\hat n_i$ that takes two values $0$ for empty site, and $1$ for occupied one.
The state of the adatom system is fully described by the set of lattice sites $\{\hat n_i\}$.
For simplicity,  we neglect possible interaction between adatoms at different sites, which can exceed $k_\text{B}T$, but should be smaller than the height of potential barrier (activation energy) separating lattice sites.
In the case when the time duration $\tau_0$ of adatom jumps from one site to other is much smaller than the average lifetime $\tau_l$, one can write the balance equation for a change in the population of the $i$-th site over time $\Delta t$ ($\Delta t\gg \tau_0$) \cite{Chumak1980,chumak1992book}:  
\begin{equation}\label{eq:balanceA}
\hat n_i(t+\Delta t) - \hat n_i(t)=\sum_j\left(I_{ji}-I_{ij}\right),
\end{equation}
where $I_{ij}$ is the number of jumps of adatoms from the $i$th to the $j$th minimum during time $\Delta t$.

For $\Delta t\ll \tau_l$, one can introduce the mean frequency of jumps $\nu_{i,j}$, and extract from $I_{i,j}$ the mean number of jumps $\nu_{i,j}\hat n_i\hat h_j\Delta t$ during the time $\Delta t$, here $\hat h_i=1-\hat n_i$ describes vacancies. The quantity $\delta I_{i,j}=I_{i,j}-\nu_{i,j}\hat n_i\hat \hat h_j\Delta t$ is the fluctuation of the jump numbers from the $i$th site to $j$th one per unit time. Equation (\ref{eq:balanceA}) can be rewritten as \cite{Chumak1980}:
\begin{equation}\label{eq:withJLA}
\left[\hat n_i(t+\Delta t)-\hat n_i(t)\right] / \Delta t=\sum_j\left(\nu_{ji} \hat n_j \hat h_i-\nu_{ij} \hat n_i \hat h_j\right)+\delta I_i,
\end{equation}
where the term $\delta I_i=\sum_j(\delta I_{j,i}-\delta I_{i,j})$ is Langevin  source of fluctuation in the evolution equation for  the population  of $i$th site. In the simplest case one can estimate the form of the correlation function of this source $\left\langle \delta I_i(t) \delta I_{i'}\left(t'\right)\right\rangle$. If we assume that the adatom jumps occur due to spatially and temporally uncorrelated impulses of substrate atoms then $\left\langle \delta I_{ij}(t) \delta I_{i'j'}(t')\left(t'\right)\right\rangle\neq 0$ only for $i=i'$, $j=j'$, $t=t'$. Taking into account that $\left\langle \delta I_{ij} \nu_{ij} n_i h_j\right\rangle=0$ (by the definition of $\delta I_{ij}$), and the probability of two or more jumps from the $i$th minimum to the $j$th within time $\Delta t$ is negligible (i.e., $I_{ij}=0$ or 1), that means \( \left\langle I_{ij}^2\right\rangle=\left\langle I_{ij}\right\rangle \), we  find
\begin{eqnarray}\label{eq:LangMeanA}
\left\langle\left(\delta I_{ij}\right)^2\right\rangle&=&(\Delta t)^{-1}\left\langle \nu_{ij} \hat n_i \hat h_j\right\rangle-\left\langle \nu_{ij}^2 \hat n_i \hat h_j\right\rangle\\
&\approx&(\Delta t)^{-1}\left\langle \nu_{ij} \hat n_i \hat h_j\right\rangle.
\end{eqnarray}

Using (\ref{eq:LangMeanA}), we obtain the  correlation function of the Langevin source of fluctuations, see \cite{Chumak1980,chumak1992book}
\begin{align}\label{eq:LangevinCorA}
\left\langle \delta I_i(t) \delta I_{i'}\left(t'\right)\right\rangle =& \nonumber \\
\delta\left(t-t'\right) \sum_j\left(\left\langle \nu_{ij} \hat n_i \hat h_j\right\rangle  + \left\langle \nu_{ji} \hat n_j \hat h_i\right\rangle\right)&\left(\delta_{ii'}-\delta_{ji'}\right).
\end{align}
The Langevin source can be considered as $\delta$-correlated only for the time scale that is much longer than $\tau_0$. Obtained equations (\ref{eq:withJLA}) and (\ref{eq:LangevinCorA}) describing fluctuation in nonequilibrium system have general form, and are not related to the specific form of statistical operator. They  are based on two different time-scales: the mean-lifetime $\tau_l$ of adatom at site, and the time $\tau_0$ of its transition to other site.

\medskip
In the same manner we can write down the equations describing fluctuations in multicomponent adsorbate system \cite{EPJST2013}:
\begin{equation}\label{balance+A}
[\hat n_i^\alpha (t+\Delta t)-\hat n_i^\alpha(t)]/\Delta t=\sum_j \bigl (\nu^\alpha_{ji} \hat n_j^\alpha \hat h_i-\nu^\alpha_{ij} \hat n_i^\alpha \hat h_j \bigr ) + \delta I_i^\alpha,
\end{equation}
where   $\hat n_i^{\alpha}=\{0,1\}$ are the site occupation numbers corresponding to the different  sorts  of adatoms $\alpha$. Here $\hat h_j=1-\sum_\alpha \hat n_j^\alpha$, and $\nu_{ij}^{\alpha}$ is mean jump frequency of $\alpha$-particle from $i$th to $j$th site.  We also assume that each site can be occupied only by one particle.    
The Langevin source of the current fluctuations $\delta I_i^\alpha$ on site $i$ reads
$\delta I_i^\alpha=\sum_j \delta I_{ji}^\alpha-\delta I_{ij}^\alpha$, where $\delta I_{ij}^\alpha=I_{ij}^\alpha-\nu^\alpha_{ij} \hat n_i^\alpha \hat h_j \Delta t$.

\medskip
Our particular interest is the case of two component lattice gas with the one ``heavy'' component, say $\alpha=0$, so that $\nu_{ij}^1\gg\nu_{ij}^0$.
For simplicity we assume that the ``heavy'' component $\hat n_i^0=\hat u_i$  is static, i.e., describes impurity adatoms, equation for mobile component $\hat n_i^1=\hat n_i$  has the same form as  Eq.~(\ref{eq:withJLA}), where $\hat h_i=1-\hat n_i-\hat u_i$.
This equation, as a result, describes the kinetics and fluctuations in the limiting case of adsorbed layer doped by impurity (heavy) adatoms.

\medskip
\textbf{2. The local-equilibrium approximation.} In the present form, Eqs.~(\ref{eq:withJLA}), and (\ref{eq:LangevinCorA}), we face into difficulties in  description of fluctuation in such system. It is to adequately introduce the fluctuations of site occupation numbers $\delta \hat n_i=\hat n_i-\langle \hat n_i\rangle$, which are not small in general.
Since the root-mean-square fluctuation of the concentration of any minimum $\left\langle \delta \hat n_i^2 \right\rangle = \left\langle \hat n_i^2 \right\rangle - \left\langle \hat n_i \right\rangle^2 = \left\langle \hat n_i \right\rangle - \left\langle \hat n_i \right\rangle^2 $, the relative magnitude $\left\langle \delta \hat n_i^2 \right\rangle / \left\langle \hat n_i \right\rangle^2 \sim 1/\left\langle \hat n_i \right\rangle$ can be quite large.
Another problem that is also related to discrete value of $\hat n_i=\{0,1\}$, is to introduce the time derivative, see, e.g., \cite{richards1977,tahir-kheli_correlated_1983}.
To avoid such difficulties it is often used a coarse-grained description of the system, see e.g., \cite{zubarev1973, Chumak1980, Ropke2013}.
The averaging is made not over the states of particles, but over a certain time interval $\tau'$, satisfying the condition $\langle\nu_{ij}\hat n_i\hat h_j \rangle \gg (\tau')^{-1}$. This means that a large number of particles visit a given site over the time $\tau'$. This allows us to consider the $\tau'$-averaged values of ${\langle\delta \hat n_i\rangle}_{\tau'}$ as small, since it is about the fluctuations of a large number of particles. After such averaging, we lose information about ``fast'' processes.

The idea of the local equilibrium approximation \cite{zubarev1973,Ropke2013} is based on the existence of two different time scales of a system relaxation to equilibrium state (the fast and slow time-variables). The first (the fast time) is characterized by the rate of the establishment of the local equilibrium at a site (or a potential relief minimum), that is of the order of $\langle\nu_{ij}\hat n_i\hat h_j\rangle^{-1}$. During this time, the establishing of the equilibrium of each site with its nearest environment occurs. Further, the process of the system relaxation goes on macroscopic spatial scales.

To describe the evolution of fluctuations ${\langle\delta\hat n_i\rangle}_{\tau'}$ for the macroscopic time-scale $t$ (the slow time) it is needed to eliminate in some way the fast processes by formal averaging of our equation over the time scale $\tau'\approx t$
\begin{equation}\label{eq:withJL_673_01A}
    \partial_t{\langle \hat n_i\rangle}_{\tau'} =
    \sum_j{\left\langle\nu_{ji} \hat n_j \hat h_i-\nu_{ij} \hat n_i \hat h_j\right\rangle}_{\tau'}+\left\langle \delta  I_i \right\rangle_{\tau'},
\end{equation}
and obtain the equations for $\delta n_i(t) = {\langle\delta \hat n_i\rangle}_{\tau'=t} \equiv\langle\delta \hat n_i\rangle$ in closed form.
This problem is related to the one of the elimination of fast variables.
For the particle-field systems, e.g., atom-photon, particle-phonon ones, the elimination of fast variable, like photon or phonon, can be made in the second order of perturbation theory that leads to the equation for the reduced atom (particle) density matrix in the Lindblad form \cite{carmichael1993open,gardiner2004,breuer2002theory,gardiner}.
The phenomenological approach based on the classical lattice gas initially faces the problem of the absence of an explicitly given field (e.g., phonons) that induces the adatom jumps between different sites (potential relief minima).
By these reasons, in order to phenomenologically describe the evolution of the system in slow time $t$, the additional assumptions are usually used.

One of the consistent ways to describe the fluctuations $\delta n_i$ and obtain Langevin equation of its evolution for the macroscopic time-scale (the slow time) is to use the approach of the local equilibrium statistical operator \cite{zubarev1973}. The averaging of physical variables by the statistical operator is equivalent to the averaging over infinite time-interval (the ergodic hypothesis).
In the local equilibrium approach \cite{Ropke2013,zubarev1973}, it is supposed that averaging over finite time scales $\tau'$ is equivalent to the averaging over statistical operator $\hat{\rho}(t)=\hat\rho^\text{e}(\mu+\delta\mu_i(t),T+\delta T_i(t))$, the form of which is similar to equilibrium statistical operator\footnote{
Here we assume that $\hat{\rho}^\text{e}=Z^{-1}\exp[(\mu \hat N-\hat H)/k_BT]$, where  $\hat N$ is the total particle number operator, $\hat H$ is the Hamiltonian given in occupation number representation.}
$\hat\rho^\text{e}(\mu,T)$, but with chemical potential $\mu(t)=\mu+\delta \mu_i(t)$ and temperature $T(t)=T+\delta T_i(t)$ slowly evolving in time and varying in space.\!\footnote{Introduction of local chemical potential is a usual practice to describe weakly inhomogeneous plasma, for example, the Debye–H\"uckel screening length \cite{ziman1972pts}.}
The contribution of $\delta T$ is usually omitted due to high heat transfer into substrate.
The averaging of Eq.~(\ref{eq:withJL_673_01A}) over $\delta\hat{\rho}(t)=\hat{\rho}(t)-\hat{\rho}^\text{e}\approx\hat\rho^\text{e}(k_\text{B}T)^{-1}\sum_i\delta\mu_i(\hat n_i-n^\text{e})$ enables one to obtain, in the explicit form, the Langevin equation and diffusive correlation function of Langevin source, describing the fluctuations $\delta n_i(t)=\Sp\{\delta \hat{\rho}(t)n_i\}$ near homogeneous or weakly-inhomogeneous equilibrium state $n^\text{e}$, see \cite{Chumak1980,chumak1992book}.

\medskip
However, we are interested in the case of fluctuations near strongly inhomogeneous nonequilibrium steady states caused by external driving field. We resort to the mean-field approximation instead of the local equilibrium operator approach.\!\footnote{We do not resort to the description in terms of local statistical operator  by yet another reason. The description in terms of chemical potential (the grand canonical ensemble) can lead to partial suppression of the local density fluctuations by particle reservoir in compared with the system with fixed particle number.}

\medskip
\textbf{3. Fluctuations near nonequilibrium steady states. The mean field approximation.}
Up to this point we do not specify the asymmetry of forward-backward particle jumps $\nu_{ij}\neq\nu_{ji}$ that is related to the form of potential relief.
In particular, the asymmetry can be caused by an external field that induces particle drag. In the case of an infinite system, driving field can be given by conservative force that leads to asymmetric particle jumps $\nu_{ji}\approx\nu[1+\bg\cdot(\br_i-\br_j)/\ell]$ where  $|\bg|=\ell|\bG|/(2kT)<1$, $\ell$ is the lattice constant, and $\bG$ is external drive, see \cite{pre2017}. This form corresponds to the one of the wind force often encountered in  electromigration  of adsorbed atoms on a solid-state substrate, see, e.g., \cite{krug1997,ishida1994,Verbruggen1988}.

To describe nonequilibrium steady states $n_i^s$, and fluctuation $\delta n_i(t)$ near them in the framework of the local equilibrium approach we apply the mean field approximation. 

We represent the site occupation numbers in the form $\hat n_i=\langle \hat n_i \rangle^0 _{\tau'=t}+\delta \hat n_i$, assuming that $\langle \hat n_i \rangle ^0 _{\tau'=t}\equiv\langle \hat n_i \rangle ^0=n_i(t)$ is  governed by the mean-field equations for the slow time variable $t$,
\begin{equation}\label{eq:8A}
  \partial_t  n_i  = \sum_j \nu_{ji} n_j h_i - \nu_{ij}n_ih_j,
\end{equation}
that is the lattice version of the mean field Smoluchowski Equation \cite{chavanis2019,schmittmann1995,leung_novel_1994,pre2017}.
Substituting $\hat n_i$ into Eq.~(\ref{eq:withJL_673_01A}) and averaging it over the  time $t=\tau'$ we obtain the Langevin equation describing fluctuations $\langle\delta \hat n_i\rangle_{t=\tau'}\equiv\langle\delta \hat n_i\rangle$ for the macroscopic- (slow-) time scales:
\begin{align}\label{eq:n_bar_nA}
 \partial_t \langle\delta \hat n_i\rangle=&
 \sum_j\Big\langle\nu_{ji} \left( n_j+\delta \hat n_j\right) ( h_i-\delta \hat n_i)\nonumber\\
    -&\nu_{ij}\left( n_i+\delta \hat n_i\right) ( h_i-\delta \hat n_i)\Big\rangle \nonumber\\
    -&\sum_j\left(\nu_{ji} n_j h_i-\nu_{ij} n_i h_j\right) + \delta \tilde I_i.
\end{align}

Our phenomenological approach based on the mean field equations (\ref{eq:8A}) and (\ref{eq:n_bar_nA}) enables us to consider the fluctuations near nonequilibriun steady states $n^s_i$ those can be obtained as solutions of
$\sum_j \nu_{ji} n_j^s h_i^s - \nu_{ij}n_i^sh_j^s=0$,
Eq.~(\ref{eq:8A}) at $\partial_t n^s_i=0$.  Taking into account that $\delta n_i\gg {\langle \delta \hat n_i \delta \hat n_j\rangle}$ we can linearize Eq.~(\ref{eq:n_bar_nA}) which is written as 
\begin{align}
 \partial_t \delta  n_i=&
 \sum_j \left[\nu_{ji} \left(h_i^s\delta n_j -n_j^s\delta  n_i\right)
    -\nu_{ij}\left( h_j^s\delta n_i-n_i^s\delta  n_j\right)\right]\nonumber\\
    &+ \delta \tilde I_i.
\end{align}
The correlation function for Langevin source associated with slow time takes the form 
\begin{multline}\label{eq:LangevinCor_2A}
\left\langle \delta \tilde I_i(t) \delta \tilde I_{i'}\left(t'\right)\right\rangle \approx
\delta\left(t-t'\right) \\
\times\sum_j \left(\nu_{ij} n^s_i h^s_j + \nu_{ji} n^s_j h^s_i\right)\left(\delta_{ii'}-\delta_{ji'}\right).
\end{multline}
In the case when interaction between particles located at different sites is negligible, one can apply the mean-field factorization $\langle n_in_j\rangle \approx \langle n_i\rangle\langle n_j\rangle$ directly  to Eq.~(\ref{eq:withJL_673_01A}),\footnote{
For interacting particles, such simple factorization is not applicable. The interparticle interaction usually exceeds $k_\text{B}T$, and the correlation between them can be  significant.}
and obtain the lattice form of the stochastic mean field Smoluchowski equation \cite{chavanis2019,schmittmann1995,leung_novel_1994,pre2017}. In the case of the representation Eq.~(\ref{eq:n_bar_nA}), the mean-field approximation $\langle \delta \hat n_i \delta \hat n_j\rangle\approx\langle \delta \hat n_i \rangle\langle \delta \hat n_j\rangle$ enable us to write down the mean field Langevin equation for macroscopic fluctuation $\delta n_i(t)=\langle \delta \hat n_i \rangle$.

Obtained phenomenological equation is valid only for long-time scales and neglects any fast processes.
In addition,  we lose the possible correlations between particles during the short times such as local memory effect, the increasing of probability of the next back jump of the particle near the particle of another sort \cite{tahir-kheli_correlated_1983}. 
We also neglect the short-range correlations that can be especially significant in the case of interparticle interaction which usually exceeds $k_\text{B}T$.

In what follows we consider the quasi-one-dimensional case corresponding to narrow channel.

\begin{figure}[t]
    \includegraphics[width=\columnwidth]{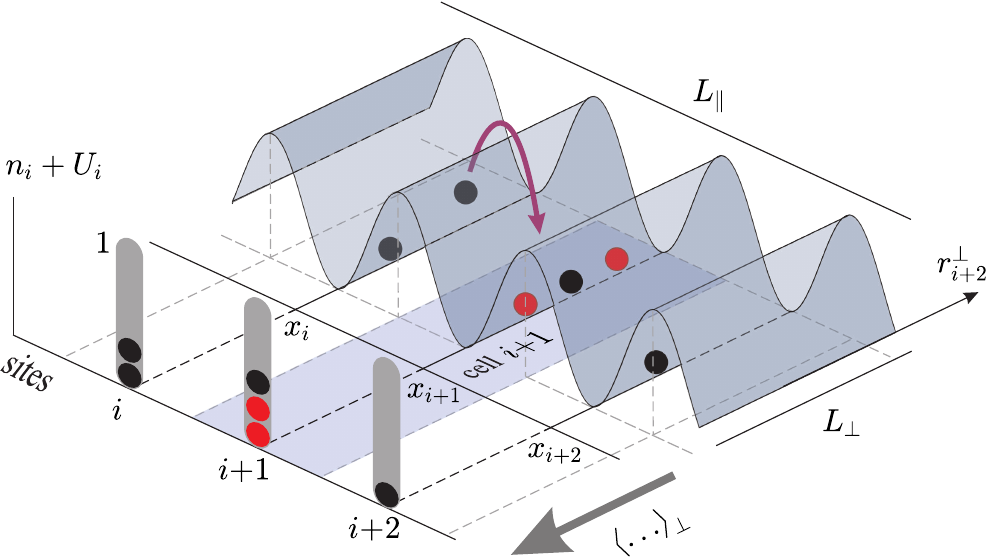}
    \caption{\label{fig:q1Dch} Schematic illustration of quasi-one-dimensional lattice gas corresponding to narrow channel system. Black circles correspond to mobile particles and red (gray) ones to impurity (``heavy'') particles. The cell $i+1$ is partially occupied by impurity particles, that corresponds to partially penetrable site $i+1$ for 1D lattice.}
\end{figure}

\medskip
\textbf{4. The quasi-one-dimensional driven lattice gas model on a ring with obstacle.}
We consider the particular case of a lattice in the form of a narrow channel with longitudinal $L_{\parallel}$ and transverse $L_{\perp}$ sizes, so that $L_{\parallel}\gg L_{\perp}$.
According to \cite{korniss_1999,benichou_2018,Miron_2020,miron2021}, the case of a narrow channel can be effectively considered as quasi-1D lattice gas.
The quasi-one-dimensional approximation for a narrow channel can be performed at any step.
This approximation is related to the additional averaging of the equation over the transverse size $L_\perp$ of a channel.
We represent the coordinate of site $i$ as $\bm{\mathrm{r}}_i = (x_i, \bm{\mathrm{r}}_i^\perp)$, where $x_i$ is the coordinate along the channel, and $\bm{\mathrm{r}}_i^\perp$ is its transverse coordinate (relatively to $\bm{\mathrm{x}}$), see Fig.~\ref{fig:q1Dch}.
Then $i$ numerates the transverse cells along the channel.
Since the averaging over transverse coordinate $\bm{\mathrm{r}}^\perp_i $ does not change the total particle number $N_i$ in $i$th cell ${\langle \hat{n}_{i,\bm{\mathrm{r}}^\perp_i}\rangle}_\perp=N_i/ L_{\perp}$, and the interaction between particles located on different sites is negligible one can use the approximate mean-field  
 factorization ${\langle\langle \hat n_i \hat n_j\rangle_t\rangle}_\perp \approx {\langle\langle\hat n_i\rangle_t\rangle}_\perp {\langle\langle\hat n_j\rangle_t\rangle}_\perp$.
As a result we obtain the mean field Smoluchowski equation, similar to Eq.~(\ref{eq:8A}), for  quasi-one-dimensional channel, where $i$ numbers the sites of quasi-1D lattice.\!\footnote{In the case $L_\parallel \gg L_\perp \gg \ell$, the averaging over transverse size may be enough to describe the system in terms of the mean site occupation numbers $0<n_k<1$ and to use the mean-field approximation without the averaging over long-time scale.}
Now, the long-time dynamics of the quasi-1D system, Eq.~(\ref{eq:8A}), is written in terms of the mean occupation numbers (concentrations) of lattice site $i$ by gas particles $n_i(t)$, corresponding to its mean concentration in $i$th cell,  vacancies $h_i=1-n_i-U_i$, and  impurities $U_i$:
\begin{equation}\label{eq:n_h_U_A}
 n_i={\langle{\langle\hat n_i\rangle}_t\rangle}_\perp, \  h_i=1-{\langle\hat u_i\rangle}_\perp-{\langle{\langle\hat n_i\rangle}_t\rangle}_\perp,\  U_i={\langle\hat u_i\rangle}_\perp,   
\end{equation}
so that $0\leq n_i(t)+U_i\leq 1$ for each site, see, e.g., \cite{schmittmann1995,leung_novel_1994,pre2017}. It is assumed that the particles of the mobile gas component perform jumps only between neighboring sites, i.e., $j=i\pm1$, and  external driving field is applied along the quasi-one-dimensional channel.
Formally, considered model can be associated with mean-field version of ASEP \cite{janowski1992,janowsky1994,costin2012,mallick2015,kolomeisky_1998,GREULICH20081972,Greulich_2008,Soh2018,Soh2017,cirillo2016,sarkar2014}.

In this setting, the gas kinetic equation (\ref{eq:8A}) for mean occupation numbers $n_i(t)$ takes the form
\begin{equation}\label{eq:nk_A}
  \dot{n}_i=J_{i-1,i}-J_{i,i+1},
\end{equation}
where $J_{i,i+1}=\nu^+ n_i h_{i+1}-\nu^- n_{i+1}h_{i}$ is bond current between $i$ and $i+1$ sites, and $\nu^{\pm}=\nu \pm \delta \nu$ are forward-backward particle hopping rates between nearest sites.
The asymmetry $\pm \delta \nu=\pm \nu g$ is caused by the action of  driving force or field $g$ which is supposed to be  nonconservative   for the chain with
periodic boundary conditions.

\textbf{5. Fluctuations induced by external drive noise.}
This phenomenological approach does not specify the stochastic field which initiates the jumps of the particles between different sites, and, as result, does non specify the form of the Langevin source in the terms of this field. Usually, the correlation function of Langevin source,  is associated with thermal fluctuations initiated by a substrate vibrations (or phonons).
In general, the jumps can be caused by particle interactions with different fields simultaneously, as is often case in solid-state state physics.\!\footnote{For instance, the interaction of adsorbed atoms (or some quasi-particle)  with phonon field of a solid-state substrate and external electromagnetic (photon) one. It is so could photo-induced diffusion realized, for example, due to the recoil effect for an atom by spontaneous photon emission, which described by atom recoil temperature,  usually low.}
In this approach we can not directly
separate the contributions of the different mechanisms into particle jumps.
We can do this only by the phenomenological constant of the mean hopping rate $\nu_{ij}$.  In our particular case, the particle jumps can by associated on the one hand, with thermal fluctuations in a substrate that determines a diffusive (advection-diffusion) processes characterized by hopping rate $\nu$, and on the other hand, with fluctuations or noise of the external driving field $g+\delta g(t)$ near its stationary value $g$.
In the case when the typical frequencies $\lambda$ of the  field noise are much smaller than ones of thermal fluctuations, i.e., $\langle \nu_{ij}\hat n_i\hat h_j\rangle_{\tau '} \gg \lambda$,  one can extract noise-induced fluctuations.
For long-time scales $t\gg\lambda^{-1}$,  the Langevin equation for such gas fluctuations $\delta n_i$ can be written in customary form where Langevin source is exactly given in terms of the stochastic field (processes) $\delta g$. In the linear approximation over small $\delta n_i$, and $\delta g$, the gas fluctuations near nonequilibrium steady state  are described by Langevin equations with additive noise
\begin{eqnarray}\label{eq:DnDgA}
    \partial_t \delta n_i &=& C_i^g \delta n_{i-1} - \left[ \bar C_{i-1}^g + C_{i+1}^g\right]
    \delta n_i + \bar C^g_i \delta n_{i+1} \nonumber\\
    &+& \delta \tilde I_i^g.
\end{eqnarray}
Here, $C_i^g = \nu^+ h_i^g + \nu^- n_i^g$, $\bar C_i^g =\nu^+ n^g_i + \nu^-h^g_i$, where $h_i^g = 1 - U_i - n_i^g$, and $n^g_i$ is nonequilibrium steady-state solution of Eq.~(\ref{eq:nk_A}) at given $g=\text{const}$. The Langevin source is written as 
\begin{equation}\label{eq:CorrDnA}
    \delta \tilde I_i^g = \nu \delta g \left[ (n_{i-1}^g - n_{i+1}^g)h_i^g + n_i^g(h^g_{i-1} - h_{i+1}^g)\right].
\end{equation}
The general form of these stochastic equations corresponds to the ones with multiplicative noise, Eq.~(\ref{eq:nk_A}). The equation (\ref{eq:nk_A}), and (\ref{eq:DnDgA}) are of our particular interest in the present work.

\textbf{6. Numerical calculation.}
For all the numerical calculations we use discrete-time version of  Eq.~(\ref{eq:nk_A}): 
\begin{equation}\label{eq:A6:nk}
  n_i(\tau_k) - n_i(\tau_k+\Delta \tau) = \Delta \tau \left( J_{i-1,i}-J_{i,i+1} \right),
\end{equation}
with  
\begin{equation}\label{eq:A6:Jk,k+1}
J_{i,i+1}=\nu^+(\tau_k) n_i(1-n_{i+1}-U_{i+1})-\nu^-(\tau_k) n_{i+1}(1-n_{i}-U_{i}),
\end{equation}
where $\nu^{\pm}=1 \pm g(\tau_k)$ and $\tau_k=\nu t_k$ is dimensionless time.
The steady-state solution of Eq.~(\ref{eq:A6:nk}) is obtained as a limit  $\tau_k\gg \Delta \tau$ (i.e., NESS at $t\rightarrow\infty$), for $\nu^{\pm}=1\pm g$, and $g=\text{const}$. In order to control the precision of steadiness, the resulting numerical NESS was regarded as finally established if $\max_k[n_k(\tau)-n_k(\tau-\Delta\tau)]\leq10^{-30}$, cf. Fig.~\ref{fig:1}. This condition of iteration procedure termination ensures that the local changes of the density profile during $\Delta\tau$ become small enough.
To describe fluctuations in a gas induced by the field noise $\delta g(\tau_k)$ we use solution of Eq.~(\ref{eq:A6:nk}), where $\nu^{\pm}=1 \pm [\bar g + \delta g(\tau_k)]$ and $\delta g(\tau_k)$ is generated random realization of a stochastic process.

\section{\label{sec:appendix} Continuum limit. Linear approximation of Burgers equation.}
\begin{figure}[h]
    \includegraphics[width=.6\columnwidth]{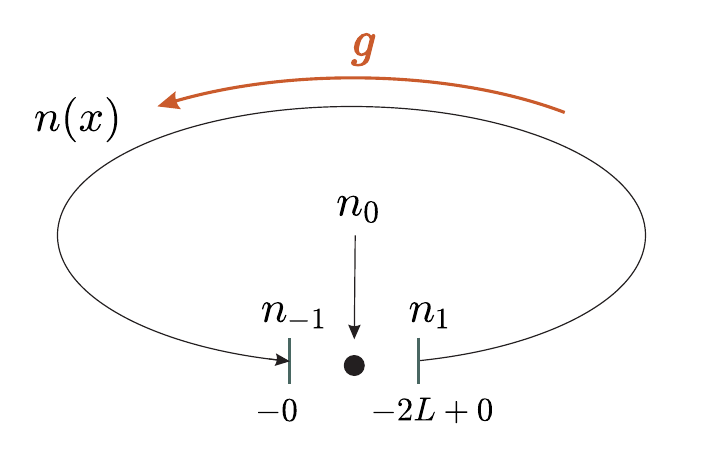}
    \caption{Continuum representation for the ring, where $n(-0)=n_{-1}$ and $n(-2L+0)=n_1$ correspond to site occupations  for lattice ring respectively.
    \label{fig:cont ring}}
\end{figure}

Here we consider continuum limit for gas concentration outside the impurity introducing continuum coordinate $k\rightarrow x$ and setting $n_k\rightarrow n(x)$, see Fig.~\ref{fig:cont ring}.
Quantity $n_{k+1}$ in this approximation is written as
\begin{equation}\label{eq:A:n-k+1}
n_{k+1}\rightarrow n(x+l)=e^{l\partial_x}n(x),
\end{equation}
where $\partial_x \equiv \partial/\partial x$.
In what follows we use dimensionless coordinate $x/l\rightarrow x$. In this representation, stationary equation outside impurity $J_{k-1,k}-J_{k,k+1}=0$ is approximated (taking into account the first nonvanishing terms) by the stationary Burgers equation
\begin{equation}\label{eq:A:burgers}
\partial_x\left[-\partial_x n + 2 g n(1-n)\right]=0,
\end{equation}
or
\begin{equation}\label{eq:A:jx=J}
-\partial_x n + 2 g n(1-n)=J=\text{const}.
\end{equation}
For convenience, we change the coordinate system by taking $-2L<x<0$, see Fig.~\ref{fig:cont ring},
and suppose that $n(-0)=n_{-1}$ and $n(-2L+0)=n_{1}$ correspond to the occupations numbers on the left and the right nearest to the impurity site ones. Equations for currents at two impurity bonds $(-1,0)$ and $(0,1)$ are no change
\begin{equation}\label{eq:J-10}
J_{-1,0}=\nu^{+} n_{-1}\left(1-n_0-U\right)-\nu^- n_0\left(1-n_{-1}\right)=J,
\end{equation}
\begin{equation}\label{eq:J01}
J_{0,1}=\nu^{+} n_0\left(1-n_{1}\right)-\nu^- n_1\left(1-n_0-U\right)=J.
\end{equation}
Equations (\ref{eq:A:jx=J}), (\ref{eq:J-10}) and (\ref{eq:J01}) are supplemented by one for the conservation of the particle number in the ring or mean gas concentration $\bar{n}$
\begin{equation}\label{eq:A:Nconservation}
\int\limits_0^{-2L}n(x)dx + n_0 = (2L+1)\bar n.
\end{equation}
For the large ring $L_0=2L+1\gg1$, we will neglect the term $n_0/(2L+1)$:
\begin{equation}\label{eq:A:Nconservation2}
  \frac{1}{L_0}\int\limits_0^{-L_0}n(x)dx \approx \bar{n}.
\end{equation}
We are interested in the subcritical case when induced by scattering gas density perturbation weakly deviates from its equilibrium value $n^e\approx \bar{n}$.   Representing gas concentration $n(x)=\bar{n}+\delta n(x)$ outside the inclusion and occupancy of impurity site  $n_0=\bar{n}+\delta n_0$ as small perturbation of mean gas concentration $\bar{n}$ one can obtain linearized form of equations (\ref{eq:A:jx=J}), (\ref{eq:J-10}), and (\ref{eq:J01}):
\begin{widetext}
\begin{subequations}\label{eq:J linear}
  \begin{equation}\label{eq:J linear a}
    -\partial_x \delta n + 2g(1-2\bar{n})\delta n=J-2 g \bar{n}(1-\bar{n})=\delta J,
  \end{equation}
  \begin{gather}\label{eq:J linear b}
    \delta n_{-1}\left[\nu^+(1-U-\bar{n})+\nu^-\bar{n}\right]-
    \delta n_0\left[\nu^+\bar{n}+\nu^-(1-\bar{n})\right]=\delta J+\nu^+U\bar{n},
  \end{gather}
  \begin{gather}\label{eq:J linear c}
   \delta n_0\left[\nu^+(1-\bar{n})+\nu^-\bar{n}\right]-
   \delta n_{1}\left[\nu^+\bar{n}+\nu^-(1-U-\bar{n})\right]=\delta J-\nu^-U\bar{n}.
  \end{gather}
\end{subequations}
Solution of Eq.~(\ref{eq:A:jx=J}) can be written as
\begin{equation}\label{eq:A:delta n}
  \delta n(x) = \frac{\delta J}{\lambda}+C e^{\lambda x},
\end{equation}
where $\lambda=2g(1-2\bar{n})$, and unknown constants $\delta J$ and $C$ can be expressed through $\delta n_{-1}$ and $\delta n_{1}$.
\begin{equation}\label{eq:A:delta n 1}
  \delta n_1 = \frac{\delta J}{\lambda}+C e^{-\lambda L_0},
\end{equation}
\begin{equation}\label{eq:A:delta n-1}
  \delta n_{-1} = \frac{\delta J}{\lambda}+C.
\end{equation}
As a result, the total system of equations to obtain unknown constants $\delta n_{-1}$, $\delta n_{0}$, $\delta n_{1}$, and $\delta J$ takes the form
\begin{equation}
    \delta n_1\left(1-(\lambda L_0)^{-1} + (\lambda L_0)^{-1} \mathrm{e}^{-\lambda L_0}\right) \approx
    \delta n_{-1}\left(-(\lambda L_0)^{-1}+\left(1+(\lambda L_0)^{-1}\right) \mathrm{e}^{-\lambda L_0}\right),
\end{equation}
\begin{equation}\label{eq:7.1b}
  \delta n_{-1}\left(\nu^{+}(1-\bar{n}+U)+\nu^- \bar{n}\right)-
  \delta n_0 \left(\nu^{+} \bar{n}+\nu^-(1-\bar{n})\right)=\delta J+\nu^{+} \bar{n} U,
\end{equation}
\begin{equation}\label{eq:7.1(c)}
  \delta n_0\left(\nu^{+}(1-\bar{n})+\nu^-\bar{n}\right)-
  \delta n_1\left(\nu^{+} \bar{n}+\nu^-(1-\bar{n}-U)\right)=\delta J-\nu^{-} \bar{n} U,
\end{equation}
\end{widetext}
\begin{equation}\label{eq:7.1(d)}
  \delta J = \lambda \frac{\delta n_1-\textrm{e}^{-\lambda L_0} \delta n_{-1}}{1-\textrm{e}^{-\lambda L_0}}.
\end{equation}
Neglecting terms of the order of $(\lambda L_0)^{-1}$, we get simple expressions for the concentration distribution $ \delta n(x)$, the current deviation $\delta J$, and site occupation numbers $\delta n_{\{-1,0,1\}}$
\begin{equation}\label{eq:A:delta n zer}
  \delta n(x) = \delta n_{-1}e^{\lambda x},
\end{equation}
\begin{equation}\label{eq:A:delta n 1 J zer}
  \delta n_1\approx \delta J \approx 0,
\end{equation}
\begin{equation}\label{eq:A:delta n 0 zer}
  \delta n_0\approx - \frac{(1-g)U\bar{n}}{1+(1-2\bar{n})g},
\end{equation}\begin{equation}\label{eq:A:delta n-1 zer}
  \delta n_{-1}\approx \frac{4}{1-U+(1-U-2\bar{n})g}\cdot \frac{(1-\nb)gU\bar{n}}{1+(1-2\bar{n})g}.
\end{equation}
In the approximation that takes into account terms of the order of $(\lambda L_0)^{-1}$, we get that $\delta n_1 \approx -(\lambda L_0)^{-1}\delta n_{-1}$, $\delta J\approx \lambda \delta n_1$, and  $\delta n(x) = \delta n_1 +(\delta n_{-1}-\delta n_1)e^{\lambda x}$.

\section{\label{sec:appendix_2}Explicit expressions for Eqs.~(\ref{eq:dispn2}) and (\ref{eq:lljllk})}

The dispersion has the form
\begin{equation}\label{eq:dispwA2}
\sqrt{\langle\delta n^2_{\langle k_s\rangle}\rangle} = {\left\langle n_{\langle k_s\rangle}^2 - \langle n_{\langle k_s\rangle}\rangle^2\right\rangle}^\frac{1}{2} = \frac{1}{2}-\langle n_1\rangle,
\end{equation}
where, substituting $\langle n_1\rangle=(n^{g-\delta g}_1 + n^{g+\delta g}_1)/2$, one obtains
\begin{equation}\label{eq:A2disp}
    \sqrt{\langle\delta n^2_{\langle k_s\rangle}\rangle} = \frac{1}{2}\left(1-(n^{g-\delta g}_1 + n^{g+\delta g}_1)\right),
\end{equation}
and $n^g_1=n_\infty(g)$ is given by Eq.~(\ref{eq:jjfkdjwi}), we get
\begin{align}
  n_1^{g\pm\delta g}=n_\infty(g\pm\delta g) \nonumber\\
  = \frac{1}{2} \left\{ 1 + \frac{n_0}{g\pm\delta g} - \left[ 1-2n_0 + \left(\frac{n_0}{g\pm\delta g}\right)^2\right]^\frac{1}{2} \right\},
\end{align}
That being inserted into Eq.~(\ref{eq:A2disp}) gives the complete expression for dispersion for arbitrary $\delta g$:
\begin{widetext}
\begin{equation}\label{eq:A2:disp_k}
     \sqrt{\langle\delta n^2_{\langle k_s\rangle}\rangle} =
     \frac{1}{4}\left(-\frac{2n_0g}{g^2-(\delta g)^2} +
     \sqrt{1-2n_0+\left(\frac{n_0}{g-\delta g}\right)^2} +
     \sqrt{1-2n_0+\left(\frac{n_0}{g+\delta g}\right)^2}
     \right).
\end{equation}
\end{widetext}
where $n_0=n_0(U)=(1-U)/2$. In the limit $g\gg\delta g$, this expression simplifies to $\sqrt{\langle \delta n_{\langle k_s\rangle}^2\rangle}\approx\tfrac{1}{2}(1-2n_\infty)$.

Characteristic enhancement of induced fluctuations of gas concentration in the vicinity of domain wall as compared to fluctuation distant from it is given by ratio Eq.~(\ref{eq:lljllk}), whose explicit form can be obtained by combining explicit expression for
\begin{widetext}
\begin{equation}
    \sqrt{\langle\delta n_1^2\rangle}=(n^{g-\delta g}-n^{g+\delta g})/2 = \cfrac{2n_0\delta g}{g^2-(\delta g)^2} + \sqrt{1-2n_0+\left(\cfrac{n_0}{g+\delta g}\right)^2} - \sqrt{1-2n_0+\left(\cfrac{n_0}{g-\delta g}\right)^2},
\end{equation}
see Eq.~(\ref{eq:dispn2}), and Eq.~(\ref{eq:A2:disp_k}):
\begin{equation}\label{eq:lljllkA2}
  \sqrt{\frac{\langle\delta n^2_{\langle k_s\rangle}\rangle}{\langle\delta n^2_1\rangle}} = \frac{-\cfrac{2n_0g}{g^2-(\delta g)^2} +
     \sqrt{1-2n_0+\left(\cfrac{n_0}{g-\delta g}\right)^2} +
     \sqrt{1-2n_0+\left(\cfrac{n_0}{g+\delta g}\right)^2}
     }{\cfrac{2n_0\delta g}{g^2-(\delta g)^2} + \sqrt{1-2n_0+\left(\cfrac{n_0}{g+\delta g}\right)^2} - \sqrt{1-2n_0+\left(\cfrac{n_0}{g-\delta g}\right)^2}}.
\end{equation}
\end{widetext}
For parameters $\delta g= 0.1$, $g=0.8$, $U=0.6$, Eq.~(\ref{eq:lljllkA2}) yields $\sqrt{\langle\delta n^2_{\langle k_s\rangle}\rangle/\langle\delta n^2_1\rangle}\approx 23.57$.
In the limit $g\gg\delta g$, the explicit expression for the enhancement ratio in Eq.~(\ref{eq:ratio}), obtained by substituting Eq.~(\ref{eq:jjfkdjwi}), has the form
\begin{equation}
    \sqrt{\frac{\langle\delta n^2_{\langle k_s\rangle}\rangle}{\langle\delta n^2_1\rangle}} \approx \frac{g^2}{n_0\delta g} \sqrt{1-2n_0+\left(\frac{n_0}{g}\right)^2}.
\end{equation}
The latter expression, for the same set of parameter values, yields $\sqrt{\langle\delta n^2_{\langle k_s\rangle}\rangle/\langle\delta n^2_1\rangle}\approx26.04$.

\end{document}